\documentclass[reqno]{amsart}
\usepackage{amssymb,latexsym,amscd}
\theoremstyle{plain}

\newtheorem{proposition}{Proposition}

\theoremstyle{definition}

\newtheorem{remark}{Remark}

\numberwithin{equation}{section}

\newcommand{\beq }{\begin{equation}}
\newcommand{\eeq }{\end{equation}}

\newcommand{\g}{\sqrt{\vert g\vert}}

\def\R{\mathbb{R}}
\begin{document}
\begin{abstract}
 In this work we analyze the relation between the  multiplicative decomposition $\mathbf F=\mathbf F^{e}\mathbf
F^{p}$ of the deformation gradient as a product of the elastic and
plastic factors (\cite{EM,Kr,L,M2}) and the theory of uniform
materials (\cite{N,TW,W}).  We prove that postulating such a
decomposition is equivalent to having a uniform material
model with two configurations - total $\phi$ and the inelastic
$\phi_{1}$. \par
  We introduce strain tensors characterizing different
types of evolutions of the material and discuss the form of
the internal energy and that of the dissipative potential.
 The evolution equations are obtained for the configurations $(\phi ,\phi_{1})$
  and the material metric $\mathbf g$.\par
   Finally the dissipative inequality for the materials of this type is presented.
It is shown that the conditions of positivity of the internal dissipation terms related to the
 processes of plastic and metric evolution provide the anisotropic yield criteria.
\end{abstract}

\title[Uniform materials and the multiplicative decomposition]%
{Uniform materials and the multiplicative decomposition of the
deformation gradient in finite elasto-plasticity}
\author{V. Ciancio}\address{Department of Mathematics,
 University of Messina, Messina, It.}\email{ciancio@unime.it}
\author{M. Dolfin}\address{Department of Mathematics,
 University of Messina, Messina, It.}\email{dolfin@dipmat.unime.it}
\author{M. Francaviglia}
\address{Department of Mathematics,  University of Torino, Torino, It.}
 \email{fviglia@dm.unito.it}
\author{S. Preston}
\address{Department of Mathematics and Statistics, Portland State
University, Portland, OR, USA} \email{serge@mth.pdx.edu} \maketitle  \today

\section{Introduction.}

The objective of this work is to investigate the relation
between the geometrical theory of uniform materials and the
multiplicative elasto-plastic decomposition of the
deformation gradient of Bilby-Kroner-Lee (BKL-decomposition) and Nemat-Nasser (see
\cite{BGS,Kr,L,M2}).\par
 Such a relation was first studied in \cite{ME}. In particular, the relation
 between the inhomogeneity velocity gradient $\mathbf L_{P}$ (see below)
and the plastic distortion rate ${\dot {\bar {\mathbf
L}}}={\dot {\bar {\mathbf F}}^p}\cdot ({\bar {\mathbf F}}^p )^{-1}
$ was introduced.  In this paper we study the geometrical form of the relation introduced in \cite{ME}.
\par

In Section 2 we introduce the basic concepts and review properties of uniform materials. In Section 3
a bijective correspondence between the BKL decompositions of the
gradient of a configuration $\phi$ of an elasto-plastic solid and the  triple $(\phi, \phi_{1},P)$ is established. Here $P$ represents the {\bf uniform material structure} and $\phi$ and $\phi_{1}$ are respectively, total and {\bf inelastic} (intermediate) material configurations.\par

In Section 4 we introduce the natural strain tensors measuring the relations between the Cauchy-Green deformation tensors $C(\phi)$ and $C(\phi_{1})$ and the material metric $\mathbf g$ induced by the uniform structure $P$. In the same section the combinations of these tensors, material metric and its curvature characteristic independent on the decomposition of plastic deformation gradient $F^p=\phi_{1*}\circ D$ are determined and the strain rate tensors are introduced.\par

 In Section 5 the form of internal energy $u$  depending on variables  $(\phi ,\phi_{1},\mathbf{g})$ and their derivatives is postulated and the dissipative potential $\mathcal{D}$ is introduced.  We also formulate the system of equations describing evolution of dynamical variables $(\phi ,\phi_{1},\mathbf{g}).$  In the same section different stress tensors present in our scheme are defined and relations between them are discussed.\par

 In section 6 we write down the dissipative inequality for the suggested scheme and separate the terms corresponding to the internal dissipation related to the processes of integrable inelastic and uniform structure evolutions.  We show that the conditions of positivity of the corresponding terms in dissipative inequality provide the anisotropic yield criteria for initiating the corresponding processes.\par

 Another form of a relation between the finite elasto-plasticity based on the multiplicative decomposition and the uniformity structures using the second-order connection was suggested by S. Cleja-Tigoui, see \cite{CT}.
\par

\section{Uniform Materials: material connections and material metrics.}

Uniform materials enter the scene of material science about 1952 when K.Kondo introduced the material connection
and the material metric as the tools to model a properties of materials. Later development
 in the works by K.Kondo, B.Bilby and his collaborators, W.Kroner, W.Noll (\cite{N}) and C.C. Wang (see \cite{TW,W}) establish the
basis of this theory.  In the works of 1980-present by M.Elzanowski,M. Epstein, M. De Leon, G. Maugin
 different aspects of this theory: models of higher grade uniform materials,
dynamics of material properties, thermodynamical properties of such materials, role of Eshelby stress
tensor, geometry of functionally graduate material, etc., were further developed.\par

In this Section we present the basic geometrical structures of the
theory of uniform materials that will be used in latter parts of the
paper. Our presentation is based on \cite{EE,EE2,EM,ME,P}.

\subsection{Material and physical spaces}

A material body ({\bf material manifold}) is usually represented by a connected $3$-dimensional
smooth oriented manifold $M$ with a piecewise smooth boundary $\partial M$. Constructions of this paper are local, so it is sufficient to consider $M$ as a connected open domain in $\R^3$ with local coordinates $X^{I}, I=1,2,3$ .\par

As the physical space our body is placed in we consider the 3-dimensional Euclidean vector space $(E^{3},\mathbf h)$, $\mathbf{h}$ being the (flat) Euclidean metric.  We introduce a global Cartesian coordinates $x^{i}$ in $\R^3$. In these coordinates the metric $\mathbf h$ takes the form $\mathbf h=h_{ij}dx^i dx^j$.\par

We will also use the concept of "archetype" (\cite{EE},\cite{M}), a 3-dimensional vector space $V$ endowed with a standard Euclidean metric and the orthonormal basis $\textbf{e}_{0}=\{e_{i},\ i=1,2,3 \}$. For convenience we identify the "archetype" space $V$ (see \cite{EM, ME}) with the tangent space at the origin $O$ of the physical space: $V=T_{O}(\R^{3})$ and its metric with the metric $\mathbf h$ at the origin. \par

\subsection{Configurations and the Cauchy metric}
\textbf{Configuration} of the body $M$ is a
(diffeomorphic) embedding $\phi : M \rightarrow E^3$ into the
physical space $E^{3}.$ To each configuration $\phi $ there
corresponds the {\bf deformation gradient} - the mapping from the tangent space $T_{X}(M)$ at the point $X\in M$
to the tangent space $T_{\phi(X)}(E^3)$ at the point $\phi(X)\in E^3$, \cite{MH},
\[ \mathbf F( X)=\phi _{*X}:T_{X}(M)\rightarrow T_{\phi (X)}(E ^{3}),\]
given, in coordinates $X^A,x^i$, by the matrix of partial derivatives
\[ \mathbf F( X)^i_I=\phi^{i}_{,I}.
\]
Here and below we will use notation $\phi^{i}_{,I}=\frac{\partial \phi ^{i}}{\partial X^{I}}( X)$
 for the partial derivatives of configuration components $\phi^i (X)$.
\par

To a configuration $\phi (X)$ there corresponds the right Cauchy-Green
deformation tensor - the flat metric $\mathbf C(\phi )=\phi
^{*}\mathbf h$ in $M$ obtained as the pullback of Euclidian metric $h$ in physical space by the configuration mapping $\phi$.
In coordinates $(X^{I})$ tensor $C(\phi)$ has the form \beq C(\phi
)_{IJ}=h_{ij}\phi ^{i}_{I}\phi ^{j}_{J}. \eeq

 We will fix a specific configuration $\phi _{o}$ and call it the {\bf reference configuration}.
Usually it presents the state of the material body that is free from
loads and stresses (see \cite{TN,M}), although it might happen that
such a configuration does not exist and one has to choose a
reference configuration differently. The body $M$ is often identified
with its image under the embedding $\phi _{o}$. \par

To the reference configuration $\phi _{o}$ there corresponds its Cauchy-Green tensor
 called the \emph{reference metric} in $M$:

\beq \mathbf g_{o}=\mathbf C(\phi _{o}),\
g_{oIJ}=h_{ij}\phi^{i}_{o,I}\phi^{j}_{o,J}, \eeq and the
corresponding \emph{reference volume form} $v_{o}(X)=\sqrt{\vert g_{o}
\vert }dX^1 \wedge \ldots \wedge dX^n .$
\par

Using the mapping inverse to the reference configuration $\phi:M\rightarrow E^3$ one can define
the frame $\mathbf p_{o}$ in $M$ by the rule
\[
\mathbf p_{o}(X)=\phi _{o,*X}^{-1}(\textbf{e}_{0}),\
(p_o)_{i}=\frac{\partial \phi^{-1\ I}_{o}}{\partial
x^{i}}\frac{\partial}{\partial X^{I}},\ i=1,2,3.
\]
\footnote{
Here and bellow for a differentiable mapping $\psi :M\rightarrow N$ between manifolds $M$ and $N$ we denote by $\psi_{*X}:T_{X}(M)\rightarrow T_{\psi(X)}(N)$ the linear mapping of tangent spaces at a point $X\in M$.  In coordinates $(X^I,x^i)$ mapping $\psi_{*X}$ is given by the matrix $F^{i}_{I}=\phi^{i}_{,I}$.  Corresponding mapping of the tangent bundles will be denoted by $\psi_{*}$: $\psi_{*}: T(M)\rightarrow T(N)$, see \cite{MH}, Ch.1.,SMK}.
\par
From now on we assume that the coordinates $X^I$ are introduced in the material manifold
$M$ using the reference configuration, i.e. $X^{I}(X)=\phi_{o}^{I}(X).$ Then the vectors of the frame $\mathbf p_{o}$ take the form $(p_o)_I=\frac{\partial }{\partial X^I},\ I=1,2,3$.\par

Finally we define {\bf a history of deformation} as a time parameterized family of smooth configurations: $\phi (t, X):
M\times\R\rightarrow E^3.$\par

\subsection{Uniform materials, I}

Recall (\cite{TW,N}) that a material is called {\bf hyperelastic} if its constitutive response (to a loading conditions) at any configuration $\phi$ is completely characterized by two scalar functions:
\begin{enumerate} \item  The {\bf elastic energy density function } (per unit of reference volume $v_{o}$) $W(X,\mathbf{F}(  X))$ depending on a material
point $ X\in M$ and the deformation gradient $\mathbf
F( X)$ at this point; and
\item
 The mass density function $\rho_{ref}(X)>0$ in the reference configuration  $\phi_{o}$.
\end{enumerate}
\par
Next we introduce the basic notion of a \textbf{uniform material (body)}.
Intuitively speaking, a \textbf{uniform body} is one that is made of the \emph{same material
at all its points}. The property of \emph{uniformity} is characterized in terms
 of a \emph{parallelism} $K^{Y}_{X}$ in the body $M$ (\cite{TW,TN,EE}). More specifically,
 a hyperelastic material body $(M,W)$ is called {\bf uniform} if for any two material points $ X,Y$ there exists a linear isomorphism $K_{ X}^{
Y} :T_{X}(M)\rightarrow T_{Y}(M)$ between tangent spaces at these points such that
\beq\label{uniformity}
 K_{ X}^{Y*}(W( Y,\mathbf{F}(Y))dv_{0}(Y))=
W( X,F(Y)\circ K^{Y}_ X )dv_{0}(X) \eeq
\newline
for all values of deformation gradients $\mathbf{F}(Y)$ at $ Y$. Here
$K_{ X}^{ Y*}$ is the pullback of the n-form of energy density by the mapping $K_{ X}^{Y}$.\par

Introduce the scalar factor $\lambda_{ X}^{ Y}$, characterizing the behavior of the reference volume form
 under the parallelism $K^{Y}_{X}$:  $K_{ X}^{ Y*}v_{o}( Y)=\lambda_{ X}^{Y}v_{o}(X).$  Then, in terms of the \emph{energy density function} $W$  condition (2.3) takes the form
\beq \label{freenergy} \lambda_{X}^{Y}W(Y,\mathbf{F}(Y))=W( X,\mathbf{F}(Y)\circ K_{X}^{Y} ) \eeq
for all points $X,Y$ in $M$ and for all values of deformation gradient $\mathbf{F}(Y)$ at the point $Y$.

\subsection{Material connections}

The localization of the definition of uniform materials given above leads to the introduction of a linear connection
({\bf material connection}) $\omega$ in $M$ having vanishing curvature (an
absolute parallelism, see (\cite{KN})). Having such a connection available, the mappings $K^{Y}_{X}$ are defined by the parallel
 translation defined by connection $\omega$ from the point $X$ to the point $Y$ along any curve connecting $X$ and $Y$ (result of
  such translation is independent on the choice of a curve due to the vanishing of the curvature). The torsion tensor $T$ of connection
$\omega$ provides the measure of \emph{non-homogeneity} of the material, see \cite{CE,EE}.
\par
It is known (see \cite{KN}, Ch.2) that in a simply connected body $M$ which admits a global tangent frame, a zero curvature
connection is determined by a choice of a \emph{global tangent frame parallel with respect to the connection} $\omega $
\[
{\mathbf p}( X)=\{ \mathbf{p}_{k}=p_{k}^{I}(X)\partial_{X^{I}},\ k=1,\ldots ,3,\ \nabla^{\omega}\mathbf{p}_{k}=0\}.
\]
\par
\begin{remark}
 A choice of such a frame is unique up to the (natural) right action
of the group $GL(n,R)$ on the tangent frames and the left action
of the {\bf symmetry gauge group} $G^M$ of the connection $\omega$
(see (\cite{EE2,P})).
\end{remark}

\par A global frame $\mathbf{p}$ may also be defined by the {\bf uniformity mapping} smoothly
depending on the point $ X$
\beq
P_{X}:V\rightarrow T_{X}(M),\ P_{X}(\textbf{e}_{i}=(P_{X})_i^{I} \partial_{X^{I}},\ i=1,2,3.
\eeq

Mapping $P_{X}$ defines the linear isomorphism of the archetype space $V$ with
the tangent space at each point $X\in M$. Section $\mathbf p$ and the uniformity map $P$ are related by
\beq
{\mathbf p}(X)=P_{X}(\textbf{e}_{0})\Leftrightarrow p_{I}(
X)=P^{J}_{I}\frac{\partial }{\partial X^J}.
\eeq

Parallel
translation $K_{ X}^{Y}$ defined by the connection $\omega$ can be
written in terms of the uniformity mapping as the composition
\[\label{parallel} K_{ X}^{{Y}}=P_{{Y}}\circ P_{X}^{-1}. \]

Using the reference frame $\mathbf p_{o}$ (see above) and the frame $\{\mathbf{e}_{i} \}$
in the space $V$, one can associate to a material frame $\mathbf{p}$ two other
geometrical objects:
\begin{enumerate}
\item A smooth mapping $k :M\rightarrow GL(V), X\rightarrow k({X})$ (an element of the gauge
group $GL(V)^{M}$) such that for all $X\in M$
\[
{\mathbf p}_{J}(X)={\mathbf p}_{o\ J}(X)\cdot k(X)\Leftrightarrow
p_{J}^{I}(X)=(p_{0\ J}^{L}k(X)_{L}^{I},\ I, J=1,2,3,
\]
here $GL(V)$ is the group of invertible linear transformations of the archetype space $V$;
\item A non-degenerate (1,1)-tensor field
$D^{I}_{J}({X})$ such that
\[
{\mathbf D}({X}){\mathbf p}_{o}( X)={\mathbf p}( X),\i.e.\
p^{I}_{i}(X)=D(X)^{I}_{J}(p_0)_{i}^{J}(X)=D^{I}_{i}(X),\
i,I=1,\ldots ,3,\]
last equality being true due to $(p_0)_{i}^{I}(X)=\delta^{I}_{i}$.
\end{enumerate}
Non-degeneracy of the (1,1)-tensor $\mathbf D(X)$ means that
$\mathbf D( X)\in GL(T_{X}(M))$.\par
Using the relation between the frame
$\mathbf p$ and the corresponding gauge mapping $ k:M\rightarrow GL(V)$ we get the relation between $k$ and the uniformity mapping $P$
corresponding to the frame $\mathbf p$, namely,
$\mathbf{p}_{i}(X)=P_{X}(\mathbf{e}_{i})=(\mathbf
p_0)_{i}k({X})=P_{o,X}(\mathbf{e}_{i})k({X}),$ so that
\[
P_{X}=P_{o,X}\circ k(X).
\]

These considerations are summarized in the following

\begin{proposition} Let $M$ be a simply connected parallelizable (i.e. admitting a global frame) manifold .
With a choice of a reference configuration $\phi_{o}$ and a frame $\textbf{e}_i $ in the archetype space $V$ there is a bijection between the following objects:
\begin{enumerate}
\item Global frames $\mathbf{p}$ in $M$ (global smooth sections of the frame bundle $F(M)$);
\item Smooth uniformity mappings $P_{X}:V\rightarrow T_{X}(M)$;
\item Smooth mappings $\mathbf k:M\rightarrow GL(V),  X\rightarrow \mathbf k(X)$ (elements of the gauge
group $GL(V)^{M}$) such that for all $ X\in M$
\[
\mathbf p( X)=\mathbf p_{o}(X) k({X});
\]
\item Non-degenerate smooth (1,1)-tensor fields $D^{I}_{J}(X)$ in $M$ such that
\[
\mathbf D( X)\mathbf p_{o}( X)=\mathbf p( X),
\]
or, in terms of uniformity mappings $P$ and $P_o$
\[
\mathbf{D}(X) = P_X \circ P^{-1}_o .
\]
\end{enumerate}
\end{proposition}
\begin{remark}
It is the bijection between the first two and the last types of
geometrical objects (non-degenerate (1,1)-tensor fields) that will
be primarily used in this paper.
\end{remark}
\subsection{Uniform materials, II}
A uniformity mapping $P$ determines its own volume form by translating to the material the Euclidian volume element from the archetype: $ v_{P}(X)=P^{-1\ *}_{ X}(\mathbf{e}_{1}\wedge \mathbf{e}_{2}\wedge\mathbf{e}_{3}).$  Denote by $J_{P}(X)$ the factor relating two volume forms $v_{o}$ and $v_{P}$
\[
v_{P}(X)=J_{P}(X)v_{0}(X)-
\]
-Jacobian of the mapping $P^{-1}$.
\par

Comparing definition of the factor $\lambda^{Y}_{X}$ in (2.4)
with the definition of the factor $J_{P}(X)$ we get, for a uniform
material following relation
between these factors: \beq \lambda_{X}^{Y}=\frac{J_{P}(
X)}{J_{P}(Y)}. \eeq

In terms of the volume factor $\mu_{P}$ uniformity condition (\ref{freenergy}) takes the form
\beq\label{freenergy1}
 J^{-1}_{P}(Y) W(Y,F(Y))=J^{-1}_{P}(X)
W(X,F(Y) \circ P_{Y}\circ P_{X}^{-1}) \eeq
\par
Combining the deformation gradient $\mathbf F(X)$ and the uniformity
mapping $P_{X}$ one gets the linear automorphism of the archetype space
$A_{X}=\mathbf{F}(X)\circ P_{X}\in GL(V)$. Comparing
(2.4) with (\ref{freenergy1}) we rewrite the condition (\ref{freenergy1}) as follows
\[J^{-1}_{P}(X)
W(X,\mathbf{F}(Y)\circ P_{Y}\circ
P_{X}^{-1})=J^{-1}_{P}(Y)W(Y,\mathbf{F}(Y))=J^{-1}_{P}(Y)W(Y,\mathbf{F}(Y)\circ
P(Y)\circ P( Y)^{-1}) \]
 for arbitrary points $X,Y\in M$ and an
arbitrary value of the deformation gradient $\mathbf{F}(Y)$ at the
point $ Y$.\par Define a function $\hat W$ of a point ${X}\in M$ and
a linear mapping  $A\in GL(V)$ by setting \beq {\hat W}(
X,A)=J^{-1}_{P}( X)W(X,A\circ P_{X}^{-1}). \eeq
\par In terms of the function $\hat W$ \emph{definition of uniform material} (\ref{freenergy1})
\emph{takes very simple form}
\beq {\hat W}( X,A)={\hat W}( Y,A). \eeq
\par Thus, {\bf the uniformity condition (\ref{freenergy}) for the
strain energy function $W$ is equivalent to the statement that the
function ${\hat W}(X,A)),\ X\in M,A\in GL(V)$ \textbf{does not depend on
the point} $X\in M$}. As a result, $\hat{W}(X,A)$ it is a function on the linear
group $GL(V)$ only.  This result is the central point of the theory
of (first grade) uniform hyperelastic materials. It reduces the
study of material properties of body $M$ and the evolution of those to the study of the uniformity mapping $P_{X}$
and the function $\hat W$ on the linear group $GL(V)$. \par

Additional physical requirements (e.g. material frame indifference, presence of a nontrivial material symmetry
group, etc.) lead to additional restrictions on the form of the energy function $W$.  For
instance, material frame indifference requirement leads to the conclusion that $\hat W (A)$ is
 a function of invariants of matrix $A$. If a uniform material is isotropic,  function $W(A)$ is left invariant
  with respect to the multiplication by elements of $SO(3)$ (\cite{TW,EE2,P}).\par

Returning to the the strain energy density function $W(X,F(X))$ we see that for a uniform
material with the uniformity mapping $P$ the strain function  $W$ takes the form (\cite{EE,EM})
\beq
W(X,F(X))=J_{P}(X){\hat W}(F(X)\circ P(X)).
\eeq

\subsection{Material metric of a uniform structure}

As it was already known to E. Cartan (see \cite{Ca}),
to a zero curvature linear connection $\omega $ (absolute parallelism)
determined by a frame $\mathbf p$ (or by the corresponding uniformity map $P$)
there corresponds the {\bf material metric} $\mathbf g$ defined as the pullback of Euclidian metric $h$ by the mapping $P^{-1}_{X}$
\beq\label{matmetric} \mathbf g({X})=P_{{X}*}^{-1}\mathbf h. \eeq\newline
This definition is equivalent to the declaring the frame $\mathbf p$ {\bf $\mathbf
g$-orthonormal} at each point $ X\in M$. In local coordinates $X^I$ the metric $\mathbf g$ has the form
\[
g_{IJ}({X})=(P_{X}^{-1})^i_I(P_{X}^{-1})^j_Jh_{ij}=(D(X)^{-1})^M_I
(D(X)^{-1})^N_J g_{0\ MN},
\]
the first expression being given in terms of the uniformity
mapping $P$ while the second is in terms of the corresponding
(1,1)-tensor field $\mathbf D$.\par
The curvature of the metric $\mathbf g$ is then defined by the torsion
of the connection $\omega $ (see \cite{MH,TN}).

\subsection{Examples}
Elastic strain tensor of a body in a configuration $\phi$ is defined by
\[
\mathbf{E}^{el}_{c}=\frac{1}{2}ln(\mathbf g_0^{-1}\mathbf C(\phi))\approx \frac{1}{2}(g_{0}^{-1}C(\phi)-I),
\]
where second expression is the linear approximation of the first one, \cite{MH,TN}.
Recall that the strain energy function of an isotropic material in linear elasticity has the form
\[
W(\phi )=\lambda [Tr (\mathbf{E}^{el})]^2+ \mu Tr[(\mathbf{E}^{el\ 2})],
\]
where $\lambda ,\mu$ are Lam\'{e} coefficients (see \cite{TN}).\par

 Using the same function ${\hat W} =\lambda [Tr(A)]^2+\mu Tr[A^2]$ on the linear group $GL(V)$ but a nontrivial uniformity mapping $P$, we come to the model of a {\bf quasi-isotropic material}.  Uniformity mapping $P$ defined the material metric $g$ as above.  This allows to \emph{redefine} the elastic strain tensor using metric $g$ instead of the reference metric $g_{0}$:
 \beq \mathbf{E}^{el}=\frac{1}{2}ln(\mathbf{g}^{-1}\mathbf C(\phi))\approx \frac{1}{2}(g_{0}^{-1}C(\phi)-I).\eeq
Strain energy of a quasi-isotropic material in linear elasticity is defied as follows

\beq W_{P}( X,\mathbf{F}({X}))=\mu_{P}({X})[\lambda (Tr(\mathbf{E}^{el}))^2+ \mu Tr(\mathbf{E}^{el\ 2})]. \eeq
It is easy to see that the strain energy is the quadratic function of the conventional elastic strain tensor $E^{el}_{c}$ with the \emph{tensor of elastic moduli depending on material point} $X$.
\par

Another example is provided by a {\bf quasi-Hookean} material (see
\cite{MH}, p.11), i.e. the uniform analog of the {\bf neo-Hookean} material with
\[
W(\phi )=\alpha [ Tr(\mathbf{E}^{el\ 2}_{c})-3].\]
 The quasi-Hookean material corresponding to a uniformity structure $P$ is defined
by the same strain energy function but with the \emph{redefined
strain tensor} $\mathbf{E}^{el}=\frac{1}{2}ln(\mathbf{g}^{-1}\mathbf C(\phi))$ where material metric $g$ is used instead of the reference metric $g_{0}$
\beq W_{P}(X,F(X))=\alpha ( Tr(\mathbf{E}^{el\ 2})-3)].\eeq
In the case of a homogeneous uniformity structure last expression reduces to
the strain energy of standard neo-Hookean material.

\subsection{Evolution of the uniform structure}

Evolution of the properties of a uniform material is characterized by the time-dependence of the uniformity mapping $P$
and that of the function $\hat W$. An appropriate characteristic of the evolution of uniform structure $P$ is the material velocity $\mathbf{L}(X)$
that was studied by different authors, see for instance \cite{ME, EM,BE}.\par

The \textbf{material velocity} of the uniformity structure $P$ is defined
as the material point and time dependent linear mapping
\[
\mathbf L_{t}(X)=P^{-1}_{X}\circ \frac{ \partial P_{X}}{\partial t}:V\rightarrow V.
\]
\par Under a loading both the uniform structure $P$ and the deformation
mapping $\phi$ are evolving.  As a result the couple $(P_{X}(t),\phi(t,X))$ (or $(g(t,X),\phi (t,X))$
describes both the (total) deformation of a material and the
evolution of its properties (elastic moduli, reference density, etc.).
The rate of change of this couple is given by $(\mathbf L_{t}(X),\mathbf{V}(t, X))$,
where $\mathbf{V}(t, X)=\frac{\partial \phi }{\partial t}$ is the physical velocity.

\section{Elasto-plastic multiplicative decompositions of the deformation gradient}

At the end of 1950s B.Bilby, E.Kroner (\cite{Kr} ) and later on E. Lee
(\cite{L}) proposed the following \textbf{multiplicative decomposition
of the deformation gradient (BKL-decomposition)}
\beq \mathbf F=\mathbf F^{e}\mathbf F^{p} \eeq as the product of
two {\bf smooth (1,1)-tensor fields} of {\bf elastic and plastic
deformations}, respectively. To provide a geometrical illustration of this decomposition an {\bf intermediate
configuration} $C^{*}_{t}$ was introduced between the material body $M$ and the current configuration $C_{t}=\phi_{t}(M)$.\par

Decomposition $\mathbf F=\mathbf F^{e}\mathbf F^{p}$ is used to study the behavior exemplified
by an elasto-plastic behavior of a material which undergoes deformation under a slowly applied load beyond the
elastic range and then, after unloading, preserves some "permanent" strain (deformation).  We refer the the monograph \cite{M2} for more examples and
 references concerning  multiplicative decompositions of the deformation gradient $F$ and their applications.

\subsection{Relation between the BKL-decomposition and the theory of uniform materials} Recall that the
deformation gradient $\mathbf F_{t}(X)$ of a configuration $\phi:M\rightarrow E^3$ is the two-point
(1,1)-tensor field in $M$ defined by the linear isomorphism
of the tangent spaces $ \phi_{*}):T(M)\rightarrow
T(\phi_{t}(M)) $ at $X\in M$.  Here $C_{t}=\phi_{t}(M)$ is the
configuration of the body at the time $t$.\par

The decomposition (3.1) can be hardly interpreted other then as the composition of tangent bundle mappings
over some mappings of corresponding base manifolds
\[
\begin{CD}
T(M) @>>> T(C^{*}_{t}) @>>> T(\phi_{t}(M))\\
@V\pi VV  @V\pi VV @V\pi VV\\
M @>>> C^{*}_{t} @>>> \phi_{t}(M)
\end{CD}
\]

since the tensor fields $\mathbf F^e$ and $\mathbf F^{p}$ should be
strictly anchored at some manifolds (domain and target of each).
Moreover the first mapping $\mathbf F^{p}$ should define a mapping
from the tangent space $T_{X}(M)$ at a point $X\in M$ to the tangent
space at some point $Y_{X}$ of intermediate configuration $ C^{*}_{t}$. The correspondence $
X\rightarrow Y_{X}$ should be one-to one, otherwise the composition
(3.1) cannot be an isomorphism of the tangent bundles. Therefore,
there exists a unique one-to one mapping $\phi_{1}:M\rightarrow
C^{*}_{t}$ underlying the tangent bundle mapping $\mathbf F^{p}$.
Mapping $\phi_{1}$ can be assumed to be differentiable.\par

In the same way $\mathbf F^{e}$ can be viewed as a mapping of tangent bundles
$T(C^{*}_{t})\rightarrow T(C_{t})$ over the differentiable mapping
$\phi_{2}:C_{t}^{*}\rightarrow T(C_{t})$ of basis manifolds.\par

We obviously have $ \phi =\phi_{2}\circ \phi_{1}$.  Therefore, $\phi_{2}$ is onto.  Restricting, if
necessary the intermediate configuration manifold one may assume, without loosing of generality, that $\phi_{1}$ is onto and
$\phi_{2}$ is one to one. Thus, both $\phi_{1}$ and $\phi_{2}$ can be
considered as diffeomorphisms.
\par

\begin{remark}  Defining the decomposition (3.1) some authors presume
that the mappings $\mathbf F^{e}$ and $\mathbf F^{p}$ are
nonsmooth or even noncontinuous, reflecting microdefects densities
in the manifold $M$. Translating this into the language of tensor
fields and using the derivatives of these tensor fields one should however assume
some smoothness. Usually it is done by considering these tensor fields as smooth \emph{averaged}
characteristics of the structural state of the material.
\end{remark}
\begin{remark} Mapping $\phi_{1}$ presents the \emph{intermediate configuration} introduced in 60th by a variety of researchers, see \cite{L,S,SB}. It was used for the construction of plastic deformation gradient $F^p$ and the elasto-plastic decompositions of total deformation gradient $F$ but as far as we know, was not considered previously as an independent dynamical variable.
\end{remark}

Now we are ready to make the next step.  Consider the tangent mapping
$\phi_{1t*}: T(M)\rightarrow T(C^{*}_{t})$ and compare it with the mapping
$\mathbf F^{p}(t,X): T(M)\rightarrow T(C^{*}_{t})$.  Since mapping $\phi_{1t*}$ is linear isomorphism at each point $X\in M$, one can write,
for all tangent vectors $\xi \in T_{X}(M)$,
\[\label{plastic}\mathbf F^{p} (t;(X,\xi ))=
\phi_{1t*X}\circ  \mathbf D_{t}(X)\cdot \xi
\]
\newline
 where $\mathbf D_{t}(X)$ is {\bf uniquely defined smooth (1,1)-tensor field  in $M$}.\par

In exactly the same way one can present
\[ \mathbf F^{e} (t,Y,\eta )= \phi_{2t*Y}\circ \mathbf
F^{e*}(t,Y)\cdot \xi \]
\newline
for the {\bf uniquely defined smooth (1,1)-tensor field $\mathbf
F^{e*}(t,Y)$ in $C^{*}_{t}$}.\par

If we pull back the (spacial index of) tensor field $\mathbf F^{e*}(t,Y)$ from
$C^{*}_{t}$ onto $M$ by the differential $\phi_{1t*}$ of the point mapping $\phi_{1}$ we get another (1,1)-tensor
$\mathbf D^{e}$ on $M$.

Since $\phi(t,X) =\phi_{t2}\circ \phi_{t1}$ for all $t$ and,
therefore, $\phi(t,X)_{*X} =\phi_{t2\ * \phi_{t1}(\mathbf X)}\circ
\phi_{t1\ *X}$, combining this with the decomposition (3.1) we get
\begin{multline*}
\phi_{*}=\mathbf F^{e}\circ \mathbf F^{p}=(\phi_{2*}\circ \mathbf
D^{e})\circ (\phi_{1*}\circ \mathbf D)=(\phi_{2*}\circ
\phi_{1*})\circ
(\phi^{-1}_{1*}\circ \mathbf D^{e}\circ \phi_{1*})\circ \mathbf D=\\
=(\phi_{2}\circ \phi_{1})_{*}\circ (\mathbf D^{e}\circ \mathbf
D)= \phi_{*}\circ (\mathbf D^{e}\circ \mathbf D),
\end{multline*}
so that
\[
\mathbf D^{e}(t,X)\cdot \mathbf D(t,X)=id_{T(M)}.
\]
As a result, {\bf being transferred to the material manifold $M$},
the (1,1)-tensor fields connecting integrable mappings
$\phi_{i*}\,(i=1,2)$ to the tangent bundles mappings $\mathbf
F^{p}$ and $\mathbf F^{e}$ are {\bf inverse to one other}. This
is hardly a surprise since in the physical literature only one of these
tensors was considered as an independent dynamical variable; see
\cite{CFR,M2}.\par

In the same way, $\phi_{2}=\phi \circ \phi_{1}^{-1}$ would be also
redundant.  As a result, the only independent dynamical variables
in this scheme are diffeomorphic embeddings $\phi,\phi_{1}$ and
the material (1,1)-tensor field $ \mathbf D$.
\par
\begin{remark} One can of course choose another triple of variables as independent dynamical quantities,
for instance one may use $(\phi_{1},\phi_{2},\mathbf D)$ if it is
preferable to deal with the elastic deformation $\phi_{2}$
explicitly.
\end{remark}
\begin{remark}
We consider here only the decomposition $\mathbf F^{e}\mathbf
F^{p}$, but the same arguments would produce a geometrical
representation of the reverse $\mathbf F^{p}\mathbf
F^{e}$-decomposition as well.
\end{remark}
\begin{remark}
Notice that the choice of an intermediate configuration
$(C^{*}_{t},\phi_{1\ t})$ participating in the decomposition
(\ref{plastic}) is far from being unique. In particular, let us
show that we may formally choose the image $C^{*}_{t}=\phi_{o}(M)$
as the intermediate configuration with $\phi_{1\ t}=\phi_{o}$ being \emph{time independent}. To do this denote by $\psi:C_{t}^{*}\rightarrow
C_{o}$ the diffeomorphism $\psi =\phi_{o}\circ \phi_{1}^{-1}$.
Transfer the tensor $\mathbf F^{e}$ to $C_{o}$ as follows: $\mathbf F^{e\ 0}=\psi_{*}(\mathbf F^{e\ *}),$
where we are using the diffeomorphism $\psi $ together with its
inverse to push forward the (1,1)-tensor $\mathbf{F^{e\ *}}$. Thus we get
the mapping of tangent bundles
\[
{\hat {\mathbf F}\mathbf F}^{e}=\chi_{t*}\circ \mathbf F^{e\
o}=\chi_{t*}\circ \psi_{t*}\circ \mathbf F^{e\ *}\circ
\psi_{t*}^{-1}.
\]
\par  Define also the diffeomorphism $\chi_{t}:C_{o}\rightarrow
C_{t}$ as $\chi_{t}=\phi \circ \phi_{o}^{-1}$.
and define the {\bf mapping of the tangent bundles} by
setting:
\[
{\hat {\mathbf F}}^{p}=\phi_{o*}\circ \mathbf F^{p\ M}.
\]
Then we have $\phi =\chi \circ \phi_{o},$ and, as is easy to check,
${\hat {\mathbf F}}^{e}\circ {\hat {\mathbf F}}^{p}=\phi_{*}$
as required.\par

As a result we get a simplified scheme of the elastic-plastic
$\mathbf F^{e}\mathbf F^{p}$-decomposition of the deformation gradient $F=\phi_{*}$ of a uniform
 material. Notice that the integrable part $\phi_{1\ t}$  of the plastic
deformation gradient $\mathbf F^{p}$ is lost in this simplified scheme.
That is why it is preferable to work with the previous scheme
where the intermediate configuration is different from the image of
the reference embedding $\phi_{o}$.
\end{remark}
\begin{remark}
 Notice that the couple $(\mathbf D,\phi_{1})$ represents
\emph{another model of evolution of the material} of the same type {\bf with the
same uniformity structure}. This model of pure inelastic evolution is related
to the model $(\phi, \phi_{1},D)$ {\bf by the elastic deformation $\phi_{2}$}.
\end{remark}
\par
If we start with a time dependent uniformity
mapping $P_{t}$ and two configurations $\phi,\phi_{1}:M\rightarrow
R^3$, then one can (reversing the arguments above) construct the "elastic deformation" $\phi_{2}=\phi \circ \phi_{1}^{-1}$ and the mappings of tangent bundles $\mathbf F^{p}:T(M)\rightarrow
T(C^{*}=Im(\phi_{1,t}),\ \mathbf
F^{e}:T(C^{*}_{t}=Im(\phi_{1,t})\rightarrow T(C_{t}=Im(\phi_{t}))$,
such that the construction above returns us to the triple $(P_{t},
\phi ,\phi_{1})$.\par

Finally, there is a freedom in the choice of the decomposition
$\mathbf F^p =\phi_{1*}\circ \mathbf D$ given by an arbitrary
diffeomorphism $ \psi \in Diff(M)$:
\beq\mathbf F^p =\phi_{1*}\circ \mathbf
D=(\phi_{1}\circ \psi^{-1} )_* \circ (\psi_* \circ \mathbf D).\eeq \par
Thus, we can introduce the following equivalence relation between the pairs
$(\phi_1 , \mathbf D)$ of the (time dependent) mappings $\phi_{1}:M\rightarrow R^n$
 and nondegenerate (1,1)-tensor fields $\mathbf{D}$ in $M$.
We say that two pairs $(\phi_1 , \mathbf D),(\chi_1 ,K)$ are
equivalent if there is a diffeomorphism $ \psi \in Diff(M)$ such
that \beq \chi_1 =\phi_1 \circ \psi^{-1} ,K=\psi_* \circ \mathbf
D.\eeq

 Collecting the considerations presented in this section we get to the following conclusions
 \begin{enumerate}
 \item  BKL-decomposition $F=F^{p}F^{e}$ of the deformation gradient $F=\phi_{*}$ of a (total) configuration $\phi_{t}:M\rightarrow E^3$ \textbf{presupposes}
 the existence of (intermediate) inelastic configuration $\phi_{1\ t}:M\rightarrow E^3$ and of the non-degenerate (1,1)-tensor field $D_{t}$ in the material space $M$ such that
 \beq
 \label{plastic1}
 \mathbf F^p = \phi_{1*}\circ
\mathbf D,\eeq

\beq  \label{elastic} \mathbf F^e = \phi_* \circ \mathbf D^{-1} \circ
\phi^{-1}_{1*}.
 \eeq
\item
Configuration $\phi_{1}$ is the mapping $\phi_{1\ t}:M\rightarrow E^3$ defining the integrable part of inelastic (plastic!)
deformation gradient $\mathbf F^{p}$,
 \item (1,1)-tensor $\mathbf D$ is equal to \[\mathbf D_{t}=\phi_{1*}^{-1}\circ \mathbf F^{p}(t,X).\]

\item Vice versa, to any triple $(\phi_t ,\mathbf F^{p}_t ,\mathbf F^{e}_t )$ consisting of two configurations and a  non-degenerate (1,1)-tensor field $D_{t}$ in $M$ there corresponds the multiplicative decomposition
 \[\phi_{*}=F=\mathbf F^{e}\circ \mathbf F^{p}\]
 of the deformation gradient $F$ of the total deformation history $\phi_t$

\item Tensor field $D_{t}$ defines the time-dependent uniform structure $P_{t}=D_{t}\circ P_{0}$ (see Proposition 1)in the material body $M$.

\item  Uniform structure $D$ determines the time dependent Riemannian metric $\mathbf g_{t} $ in the material
manifold $M$ (see Sec.2.6) by the formula:
\beq
g_{t\ IJ}=h_{ij}D^{i}_{I}D^{j}_{J}.
\eeq
\end{enumerate}

\begin{remark}
Notice that the decomposition $\mathbf F=\phi_{*}=\mathbf
F^{e}\circ \mathbf F^{p}$ determines the tensor field $\mathbf D$
and the plastic integrable deformation $\phi_{1}$ up to an action of a diffeomorphism $\chi$.\par
On the other hand there are arguments showing that the inelastic configuration $\phi_{1}$ is defined uniquely by the history of deformation. If total deformation $\phi_{0}\rightarrow \phi_{t}$ is going under certain conditions of loading, heating etc, the unloading or turning of the heat at some moment $t_{1}$ produces certain configuration $\phi_{1}:M\rightarrow E^3$. Often
the unloading happens fast and is not accompanied by an essential change
of the material structure. See \cite{LA} where relaxation of a material to the intermediate configuration $\phi_{1}$ during unloading is discussed.  As a result we may associate with the moment $t_{1}$ the final configuration
 $\phi_{1\ t_{1}}$ taken by the body $M$ after unloading.
As a result, \emph{under physically reasonable assumptions on the evolution
process}, the intermediate configuration $\phi_{1}$, and therefore
the tensor field $\mathbf D$, are \emph{determined uniquely} (up to
a composition of $\phi_{1}$ with the Euclidean motion of the
physical space $\mathbf E^3$).\par
 Using the terminology of G.Maugin and
W. Muschik (\cite{MM}), the configuration  $\phi_{1}$ is "\emph{observable but not controllable}"
(or, citing the same work, "\emph{partly controllable}" through the loading conditions).
\end{remark}

In this analysis of the kinematics of the BKL-decomposition we
identified three variables: the uniform structure $P_{t}$, the
integrable part of the "plastic deformation gradient" $\phi_{1}$ (intermediate configuration),
and the total deformation $\phi$, which we will consider as {\bf
physically independent}.  Thus, the full dynamical/thermodynamical picture with this
kinematics should include all three components. Below, discussing the dynamical structure
 of the presented model, we will be using material metric $g$ instead of the tensor field $D$ as the material dynamical variable.
 Metric $g$ contains the essential information about the uniformity structure $P_{t}$ and is more convenient to use when describing the
 evolution of material properties then the tensor $D$ or uniformity mapping $P$, see \cite{EE2}.

\section{4-Metric Theory and Strain Tensors}

\subsection{Strain Tensors}
Nonlinear Elasticity Theory has, as its geometrical
keystone, the question of the comparison of two metrics: the
reference metric $\mathbf g_{o}$ and the Cauchy metric $\mathbf
C(\phi_{t})$ of a configuration $\phi_{t}$,  \cite{MH}.\par
 As Theorem 1 shows, the multiplicative decomposition
leads, in its geometrical form, to the presence of four metrics in
the material manifold $M$: the reference metric $\mathbf g_{o}$,
the material metric $\mathbf g$ generated by the (1,1)-tensor field
$\mathbf D=\mathbf F^{p\ M}$, the Cauchy metric of the integrable
part of plastic deformation $\mathbf C(\phi_{1})$ and finally, the
Cauchy metric of the total deformation $\mathbf C(\phi)$.  It seems
natural to define appropriate strain tensors as measures of comparison between pairs of these metrics
and use these tensors for description of different processes developing in the material.\par

We will introduce six strain tensors as suited to describe the state of our solid and characterize
specific processes undergoing in the body.

\begin{enumerate}

\item  {\bf Elastic strain tensor}
\[
\mathbf E^{el}=\frac{1}{2}ln[\mathbf C(\phi_{1})^{-1}\mathbf
C(\phi)]\approx \frac{1}{2}\mathbf C(\phi_{1})^{-1}[\mathbf
C(\phi)-\mathbf C(\phi_{1})]\approx \frac{1}{2}\mathbf
g^{-1}[\mathbf C(\phi)-\mathbf C(\phi_{1})].
\]
Elastic strain tensor measures the elastic part of the deformation at each instant of time and
vanishes under unloading. Tensor $E^{el}$ and its linearized variant are $C(\phi_{1}$-symmetrical (i.e. $E^{el}_{IJ}={C(\phi_{1})}_{IK}E^{el\ K}_{J}$ is symmetrical (0,2)-tensor.

\item {\bf Inelastic strain tensor}

\[
\mathbf E^{in}=\frac{1}{2}ln[\mathbf g^{-1}\mathbf
C(\phi_{1})]\approx \frac{1}{2}\mathbf g^{-1}[\mathbf
C(\phi)-\mathbf g]\approx \frac{1}{2}\mathbf g_{o}^{-1}[\mathbf
C(\phi)-\mathbf g].
\]
Inelastic (plastic) strain tensor measures the plastic but
still Euclidean {\bf deformation} of the body, i.e. permanent
after unloading but not leading to any {\bf residual stresses}
in the material.  Tensor $E^{in}$ and its linearized version $E^{in}_{lin}$ are $g$-symmetrical, i.e. (0,2)-tensors $g_{IK}E^{in\ K}_{J}, g_{IK}E^{in\ K}_{lin\ J}$ are symmetrical.

\item {\bf Material strain tensor}
\[
\mathbf E^{m}= \frac{1}{2}ln(\mathbf g_{o}^{-1}\mathbf g)\approx
\frac{1}{2}\mathbf g_{o}^{-1}(\mathbf g-\mathbf g_{o}).
\]
Material strain tensor measures the pure metrical evolution of the material, not leading to any
deformation (material points displacement). Tensor $E^m$ and its linearized version $E^{m}_{lin}$ are $g_{0}$-symmetrical.
\item {\bf Euclidian strain tensor}
\[
\mathbf E^{eucl}:=\frac{1}{2}ln[\mathbf g^{-1}\mathbf
C(\phi)]\approx \frac{1}{2}\mathbf g^{-1}[\mathbf C(\phi)-\mathbf
g].
\]
Euclidian strain tensor measures the integrable part of the
total deformation.
\item {\bf Total strain tensor}
\[
\mathbf E^{tot}=\frac{1}{2}ln(\mathbf g_{o}^{-1}\mathbf
C(\phi))\approx \frac{1}{2}\mathbf g_{o}^{-1}(\mathbf
C(\phi)-\mathbf g_{o}).
\]
Total strain tensor measures the decline of the Cauchy metric of
total deformation $\phi$ from the reference (euclidian) metric
$\mathbf g_{o}$. It is observable.  Tensor $E^{tot}$ and its linearized version $E^{tot}_{lin}$ are $g_{0}$-symmetrical.
\item {\bf Total inelastic strain tensor}
\[
\mathbf E^{tin}=\frac{1}{2}ln(\mathbf g_{o}^{-1}\mathbf
C(\phi_{1}))\approx \frac{1}{2}\mathbf g_{o}^{-1}(\mathbf
C(\phi_{1})-\mathbf g_{o}).
\]
Total inelastic strain tensor measures the decline of the Cauchy metric of inelastic deformation
$\phi_{1}$ from the reference (euclidian) metric $\mathbf g_{o}$. It is observable (after unload).
\end{enumerate}
In each case we provide the linear approximation form(s) of the strain tensors suited for small deviation of the former metric from
the latter one.
\par

\begin{remark}
In some simple cases, say when the (1,1)-tensors $\mathbf
g^{-1}\mathbf C(\phi_{1})$ and $\mathbf C(\phi_{1})^{-1}\mathbf
C(\phi)$ commute, from the relation $ \mathbf g^{-1}\mathbf
C(\phi)=\mathbf g^{-1}\mathbf C(\phi_{1})\cdot \mathbf
C(\phi_{1})^{-1}\mathbf C(\phi)$ we conclude that the linearized
{\bf Euclidean strain tensor} $\mathbf E^{eucl}$ splits as follows
\begin{multline}
\mathbf E^{eucl}\approx \frac{1}{2}\mathbf g^{-1}[\mathbf
C(\phi)-\mathbf{g}]=\frac{1}{2}\mathbf g^{-1}[\mathbf
C(\phi)-\mathbf C(\phi_{1})]+ \frac{1}{2}\mathbf g^{-1}[\mathbf C(\phi_1 )-\mathbf g]\approx \\
\approx \mathbf g^{-1}\mathbf C(\phi_{1})\mathbf E^{el}+\mathbf
E^{m}\approx \mathbf E^{el}+\mathbf E^{in}.
\end{multline}
 \par

If all three (1,1)-tensors $\mathbf g_{o}^{-1}\mathbf g$, $
\mathbf g^{-1}\mathbf C(\phi_{1})$ and $ \mathbf
C(\phi_{1})^{-1}\mathbf C(\phi)$ commute we have an additive
decomposition of the {\bf linearized total strain tensor} $\mathbf
E^{tot}$
\begin{multline}
\mathbf E^{tot}\approx \frac{1}{2}\mathbf g^{-1}_{0}[\mathbf
C(\phi)-\mathbf g_{o}]= \frac{1}{2}\mathbf g^{-1}_{0}[\mathbf
C(\phi)-\mathbf C(\phi_{1})+\mathbf C(\phi_{1})-\mathbf g+\mathbf
g-\mathbf g_{o}]=
\\ =
\frac{1}{2}\mathbf g^{-1}_{0}\mathbf C(\phi_{1})\mathbf
C(\phi_{1})^{-1}[\mathbf C(\phi)-\mathbf C(\phi_{1})]+
\frac{1}{2}\mathbf g^{-1}_{0}\mathbf{g}\mathbf g^{-1}[\mathbf C(\phi_{1})-\mathbf g]+\frac{1}{2}\mathbf g^{-1}_{0}(\mathbf g-\mathbf g_{o})= \\
= \mathbf E^{m}+\mathbf g^{-1}_{0}\mathbf g\mathbf
E^{in}+\mathbf g^{-1}_{0}\mathbf C(\phi_{1})\mathbf E^{el}\approx
\mathbf E^{m}+\mathbf E^{in}+\mathbf E^{el}.
\end{multline}
\end{remark}

\subsection{Choice of the dynamical variables}
In order to determine which combinations of dynamical variables $(\phi, \phi_{1},D)$
and their derivatives might enter the internal or free energy, dissipative potential,
entropy and other dynamical and thermodynamical quantities we have to take into account
requirements of invariance or covariance of these quantities with respect to the
appropriate material and spacial transformations.  For instance, the frame indifference requirement (\cite{Mus, Mu}) leads to the conclusion that the deformation gradient $F(X)$ of the total deformation $\phi$ enters these quantities only in combinations $C(\phi_{1})^{-1}C(\phi),\ g_{0}^{-1}C(\phi)$ or $g^{-1}C(\phi ).$\par

Material metric $\mathbf g$ and the Cauchy-Green tensor
$\mathbf C(\phi_1 )$ depend on the choice of the plastic
decomposition (\ref{plastic1}). Thus, it is important to determine which tensors or combinations of
tensors constructed from the dynamical fields $(\phi,
\phi_{1},\mathbf D)$ are independent of the
choice of the plastic decomposition (\ref{plastic1}).\par
We consider several such combinations.\par
1. The total strain tensor \beq \mathbf
E^{tot}=\frac{1}{2}ln(\mathbf g_{o}^{-1}\phi^{*}\mathbf h)\approx
\frac{1}{2}\mathbf g_{o}^{-1}[\mathbf C(\phi)-\mathbf g_{o}]\eeq
is independent of the plastic decomposition (\ref{plastic1}).
Since it is a (1,1)-tensor its invariants are independent of the
decomposition (\ref{plastic1}) and, moreover, one can combine it with other
similar tensors to produce new invariant combinations.\par

 2. The plastic deformation gradient $\mathbf F^p =\phi_{1*}\circ \mathbf D$  is invariant under
   decomposition (\ref{plastic1}) but it is convenient to transform it into a material tensor. For instance, one can use
  the following variant of the Cauchy-Green tensor
  \beq \mathbf F^{p*}\mathbf h\equiv \mathbf D^* (\phi^{*}_{1}\mathbf h )= \mathbf D^* \mathbf C(\phi_{ 1}).\eeq

Lifting one index in this tensor by means of the reference metric $\mathbf
g_0$ we get the material (1,1)-tensor $\mathbf g_{o}^{-1\ *}
\mathbf F^{p*}\mathbf h.$ This tensor carries information on both
integrable and nonintegrable parts of the plastic deformations.

3. The material metric $\mathbf g=P^{-1*}\mathbf h $ can be
written in a number of different ways.  For instance, by using the
reference metric $\mathbf g_o =P^{-1*}_o \mathbf h$ we can get
\beq \mathbf g=P^{-1*}\mathbf h=P^{-1*}(P^{* }_o \mathbf g_o)=(P_o
\circ P^{-1} )^* \mathbf g_o =(P \circ P^{-1}_o )^{-1\ * } \mathbf
g_o= \mathbf D^{-1\ * }\mathbf g_o. \eeq In local coordinates we
have
\beq g _{ AB}=(D^{-1})^{ M}_A  (D^{-1})^{ N}_B g_{o\ MN },\ g^{-1 \  AB}=D^{ A}_M  D^{ B}_N g_{0
}^{MN }. \eeq

Under the change of decomposition (\ref{plastic1}), the material
metric $\mathbf g$ is transformed by a diffeomorphism $\psi$ into
a new metric $\mathbf g^{\psi } $ as follows
\beq \mathbf g^{\psi } = (\psi_* \circ \mathbf D)^{-1\ *}\mathbf
g_0  = (\mathbf D^{-1 }\circ \psi_* )^* \mathbf g_0 = \psi^{ -1\
*}\circ \mathbf D^{-1\
* } \mathbf g_0 =\psi^{*}\mathbf g.
 \eeq
In coordinates
\[
(g^{\psi } )_{ AB}=(\psi^{-1})^{ M}_A  (\psi^{-1})^{ N}_B g_{MN },\ (g^{\psi \ -1} )^{ CD}=\psi^{ C}_M
\psi^{ D}_N g^{-1\ MN }.
\] \par
Therefore, the curvature tensor $R^I _{JKL}$ (as well as the
corresponding Ricci tensor $R_{IJ}(\mathbf g )$) is transformed
tensorially by $\psi^*$ or $\psi_*$. Thus, invariants of these
curvature tensors (in particular the scalar curvature $R(\mathbf g
)$), do not depend on a choice of plastic decomposition (\ref{plastic1}).

4. At the same time the Cauchy metric of the inelastic deformation
$\mathbf C(\phi_1 )$ is transformed by $\psi^{-1}$ tensorially as
well: \beq C(\phi_{1}\circ \psi^{-1} )_{AB} = (\psi^{-1\ *}\circ
\phi^{* }_1 h)_{AB } = [\psi^{-1\ *}C(\phi_1 )]_{AB
}=(\psi^{-1})^{ M}_A (\psi^{-1})^{ N}_B C(\phi_1 )_{MN }.\eeq \par
Combining the material metric and the Cauchy tensor of the
inelastic deformation $\phi_1$ we finally get the inelastic strain
tensor $\mathbf E^{in}.$ For a different decomposition
(\ref{plastic1}) we have
\begin{multline}
(g^{\psi \ -1} )^{ AC} C(\phi_{1}\circ \psi^{-1} )_{CB}=\psi^{
A}_M \psi^{ C}_N g^{-1\ MN}(\psi^{-1})^{ K}_C (\psi^{-1})^{ L}_B
C(\phi_1 )_{KL }=\\ =\psi^{ A}_M g^{-1\ MN}\delta^K _N
(\psi^{-1})^{ L}_B C(\phi_1 )_{KL }=\psi^{ A}_M (\psi^{-1})^{ L}_B
g^{-1\ MK} C(\phi_1 )_{KL }=\\
=\psi^{ A}_M (\psi^{-1})^{ L}_B
(g^{-1} C(\phi_1 ))^{M }_{L }.
\end{multline}

Thus, $\mathbf E^{in}$ transforms tensorially under a change of plastic decomposition (\ref{plastic1}) and its
invariants are independent of this decomposition.\par

The presence of the two quantities $\mathbf F^p$ and $\mathbf
E^{in}$ whose invariants are independent of the choice of plastic
decomposition makes it important to compare these two quantities.
We have
\begin{multline*}  \mathbf{g}_{o}^{-1}(\mathbf{F}^{p*})\mathbf{h})=\mathbf{g}_{o}^{-1}(\mathbf{D}^{*}\mathbf{C}(\phi_{1}))=g^{- AC}_{o}D^{M}_{C}D^{N}_{B}C(\phi_{1})_{MN}=   \\
D^{-1\ A}_{L}( g_{o}^{-1\
KC}D^{L}_{K}D^{M}_{C})C(\phi_{1})_{MN}D^{N}_{B}=D^{-1\
A}_{L}(g^{-1\ ML})C(\phi_{1})_{MN}D^{N}_{B}=\\ = D^{-1\
A}_{L}(g^{-1}C(\phi_{1}))^{L}_{N}D^{N}_{B}=D^{-1\
A}_{L}(exp(2E^{in}))^{L}_{N}D^{N}_{B},
\end{multline*}
i.e. \beq \mathbf g_{o}^{-1}(\mathbf F^{p*}\mathbf h)=D^{-1\
A}_{L}[exp({2E^{in}})]^{L}_{N}D^{N}_{B}, \eeq and therefore
invariants of the tensor $\mathbf g_{o}^{-1}(\mathbf F^{p*}\mathbf h)$ (i.e. tensorial characteristics of the
plastic gradient deformation $ \mathbf F^{p}$) contain the same
information as the invariants of $\mathbf E^{in}$.\par

Thus in the restricted case it seems natural to choose an internal energy $u$ as a function of the invariants of the two strain
tensors $\mathbf E^{tot}$ and $\mathbf E^{in}$, of the invariants
of the curvature tensor $R^{i}_{jkl}(\mathbf g)$ (of the Ricci
Tensor $Ric(\mathbf g)^{I}_{J}$ in the 3-dim case, see \cite{CP}),
temperature and its $\mathbf g_{o}$-gradient: $ u=u([\mathbf E^{tot}(\phi ),\mathbf E^{in}(\phi_{1},\mathbf g),
Ric(\mathbf g),\theta, \nabla^{\mathbf g_{o}}\theta ]). $\par

If we adopt the assumptions of Remark 5 (i.e. removing the
restriction to use only tensors invariant under the plastic
decomposition (3.6)) we may consider all three strain tensors
$\mathbf E^{tot},\mathbf E^{in},\mathbf E^{m}$ (one can replace
$\mathbf E^{tot}$ in this list by $\mathbf E^{el}$ if it is preferable)
 as independent dynamical variables and, together with
the Ricci tensor of the material metric $\mathbf g$, include them as
arguments in the internal energy: \beq\label{freenergy2}
 u=u[\mathbf E^{el},\mathbf E^{in},\mathbf E^{m}, Ric(\mathbf g),\theta, \nabla^{\mathbf g_{o}}\theta ].
\eeq
In this approach the effects of different types of processes are directly separated.

\subsection{Additional strain decompositions}

Between the strain tensors introduced
above,  conventional strain tensors and both deformation
gradients $F^e ,\ F^p$ of the multiplicative decomposition there exist different relations
that may be in some partial cases more convenient than those
presented above.  Below are two examples of such relations, the
first being valid in the linear case, the second in a nonlinear situation.

 1. Linear case: \par

Since $F=\phi_{*}=\mathbf F^e \circ \mathbf F^p $ we have
\[
\mathbf g_{o}^{-1}\phi^{*}\mathbf h=\mathbf g_{o}^{-1}\mathbf
F^{p*}(\mathbf F^{e*}\mathbf h-\mathbf h+\mathbf h)=\mathbf
g_{o}^{-1}\mathbf F^{p*}\mathbf h+\mathbf g_{o}^{-1}\mathbf
F^{p*}(\mathbf F^{e*}\mathbf h-\mathbf h)
\]
and therefore
\beq
\mathbf E^{tot}_{lin}=\frac{1}{2}(\mathbf g_{o}^{-1}\phi^{*}\mathbf
h-\boldsymbol\delta)=\frac{1}{2}(\mathbf g_{o}^{-1}\mathbf
F^{p*}\mathbf h-\boldsymbol\delta)+\mathbf g_{o}^{-1}\mathbf F^{p*}
\frac{1}{2}(\mathbf F^{e*}\mathbf h-\mathbf h)\approx \mathbf
D^{-1}\mathbf E^{in}_{lin}\mathbf D+\mathbf g_{o}^{-1}\mathbf
F^{p*}\mathbf E^{el}_{lin},
\eeq
where we defined
\[
\mathbf E^{el}_{lin}=\frac{1}{2}(\mathbf F^{e*}\mathbf h-\mathbf h).
\]
This decomposition can be compared with those in Section 4.1.

2. Nonlinear case:
\begin{multline} \mathbf E^{tot}=\frac{1}{2}ln(\mathbf g_{o}^{-1}\phi^{*}\mathbf h)=\frac{1}{2}ln(\mathbf g_{o}^{-1}
(\mathbf F^{p*}\mathbf h) \mathbf h^{-1}(\mathbf  F^{e*}\mathbf h ))=
\frac{1}{2}ln(\mathbf g_{o}^{-1}(\mathbf F^{p*}\mathbf h)\cdot exp(2\mathbf E^{el}_{old}))=\\
\frac{1}{2}ln(\mathbf g_{o}^{-1}(\mathbf F^{p*}\mathbf h)\cdot (
\mathbf F^{p*}exp(2\mathbf E^{el}_{old})))\approx
\frac{1}{2}ln(\mathbf g_{o}^{-1}\mathbf F^{p*}\mathbf h)+\mathbf
g_{o}^{-1}\mathbf F^{p*}(\mathbf E^{el}_{old}),
\end{multline}
here $\mathbf E^{el}_{old}=\frac{1}{2}ln (\mathbf h^{-1}\mathbf
F^{el*}\mathbf h)$ as in conventional finite elasticity.

\subsection{Strain Rate Tensors}

We define the strain rate
tensors as time derivatives of strain tensors. As a result we
get the \emph{strain rate tensors} $ \dot {\mathbf E}^{tot}, \dot {\mathbf
E}^{in},\dot {\mathbf E}^{el}$ and $ \dot {\mathbf E}^{m}$.

On the other hand there are other rate characteristics for each of the three participating structures,
i.e.:
\begin{enumerate}
\item
The material velocity:
\[
\mathbf L_{D}(t,X)=\mathbf D(t,X)^{-1}\circ \frac{\partial}{\partial
t}\mathbf D(t,X).
\]
This velocity is related to the velocity $\mathbf L_{P}=P^{-1}\circ
P_{,t}$ introduced in Sec.2.8 by the relation $\mathbf L_{D}=P\circ
\mathbf L_{P}\circ P^{-1}$. We see now that $\mathbf{{\dot D}}=\mathbf
D\mathbf L_{D}$ and therefore we get the relation between the
material strain rate tensor and the material velocity $\mathbf
L_{D}$ used in \cite{BE,M,ME} and other papers.
\begin{multline} \dot {\mathbf E}^{m}=\frac{1}{2}\mathbf g_{o}^{-1}{\dot {\mathbf g}}=
\frac{1}{2}\mathbf g_{o}^{-1}({\dot D}^{M}_{I}D^{N}_{J}g_{0\ MN}+{
D}^{M}_{I}{\dot D}^{N}_{J}g_{o\ MN})=\\ =
\frac{1}{2}\mathbf g_{o}^{-1}(D^{M}_{K}L^{K}_{D\ I}D^{N}_{J}g_{o\ MN}+ D^{M}_{I}D^{N}_{S}L^{S}_{D\ J}g_{o\ MN})=\\
=\frac{1}{2}\mathbf g_{o}^{-1}\left( L^{K}_{D\ I}g_{P\ KJ}+
L^{S}_{D\ J}g_{IS}\right)=\mathbf g_{o}^{-1}(L_{D})^{K}_{ (I}\cdot
g_{\vert K\vert J)},
\end{multline}
where in the last formula there is symmetrization by indices $IJ$.
\item
The total velocity is defined as
\[
\mathbf V(X,t)=\frac{\partial }{\partial t}\phi ( X,t)
\]
and its gradient is related to the linearized total strain rate tensor ${\mathbf E}^{tot}_{lin}=\frac{1}{2}g_{0}^{-1}(C(\phi)-g_{0})$ by the relation
\[
(g_{0}{\dot {\mathbf E}}^{tot}_{lin})_{MN}=h_{ij}V^{i}_{(,M}\phi^{j}_{N)}.
\]
Being written in Euler (spacial) coordinates this relation reduces to the standard one (\cite{TN}).
\item
Finally, the velocity of the inelastic deformation
\[
\mathbf V_{1}(X,t)=\frac{\partial }{\partial t}\phi_{1}( X,t),
\]
 is related, in a linear approximation, to the (linearized) inelastic strain rate tensor  $\dot {\mathbf E}^{in}_{lin}$ by a relation containing the symmetrized velocity gradient and the material velocity $\mathbf L_{D}$.  In the calculation that follows we are using the formula $\frac{\partial}{\partial t}C(\phi_{1})=2h_{ij}V^{i}_{1,(M}\phi^{j}_{1,N)}$ for time derivative of Cauchy-Green tensor of configuration $\phi_{1}$.  We have
\begin{multline}
{\dot {\mathbf E}}^{in}_{lin}=\frac{\partial}{\partial t} \left(
\frac{1}{2}\mathbf g^{-1}(\mathbf C(\phi_{1})-\mathbf g)\right)=
\frac{1}{2}\frac{\partial}{\partial t} \left( \mathbf g^{-1}\cdot
\mathbf C(\phi_{1})
\right)=\\
=\frac{1}{2}\mathbf g^{-1}\frac{\partial}{\partial t}\mathbf C(\phi_{1})+\frac{1}{2}\frac{\partial}{\partial t}({\mathbf g}^{-1})\cdot \mathbf C(\phi_{1})=
{\mathbf g}^{-1}[ h_{ij}V^{i}_{1,(M}\phi^{j}_{1,N)}]
+\frac{1}{2}\mathbf g^{-1}{\dot
{\mathbf g}}\mathbf g^{-1}\cdot \mathbf C(\phi_{1})=\\
=\mathbf g^{-1}[ h_{ij}V^{i}_{1,(M}\phi^{j}_{1,N)}]+ \mathbf g^{-1}(\mathbf g_{o}{\dot {\mathbf
E}}^{m})\mathbf g^{-1}\mathbf C(\phi_{1})\approx
 (\nabla^{\mathbf g_{o}}\mathbf{v}_{1})_{sym}+ {\mathbf{\dot{E}}}^m .
\end{multline}
Here $\mathbf{v_{1}}_{M}= \phi^{i}_{,M}h_{ij}\mathbf{V}^{j}_{1}$ is the covariant form of the \emph{convective velocity} of inelastic configuration $\phi_{1}$, see \cite{SMK}.  In the last approximation we replaced $\mathbf{g}\approx \mathbf g_{o}$
in the first and the second terms and $ \mathbf C(\phi_{1})\approx \mathbf g$ in the
second term.
\end{enumerate}

\section{Lagrangian, free energy, dissipative potential and the stress tensors}

Dynamical equations describing the evolution of the system
characterized by the variables $(\phi, \phi_{1},\mathbf{g})$ are
obtained by combining the canonical (Lagrangian) component and the the dissipative
forces.\par The Lagrangian in our model
is the combination of kinetic, potential and internal energy terms:
\beq L=K-\rho_{ref}u-U(\phi), \eeq where $K=\frac{\rho}{2}\vert
\mathbf{V}\vert_{h}$ is the density of kinetic energy, $U(\phi)$ is
the potential of the volume forces and $u$ is the internal energy per unit of mass (see 4.2).
\subsection{Lagrangian and internal energy}
\par
It is traditional to define free energy density $\psi$ as a function of the elastic deformation
gradient, the temperature $\vartheta $, a material point $X$ and additional internal parameters $\boldsymbol\alpha$ (see
\cite{M1,MM}, Sect. 10.1A).\par
{\bf Dissipative pseudo-potential} is, in this approach, the function of rates of deformation
 gradients and time derivatives of internal variables $\mathcal{D}=\bar D (\dot {\mathbf
F}^{e}, \dot {\mathbf F}^{p},  \mathbf{ \dot{\boldsymbol\alpha}}
,\nabla\vartheta ,\nabla {\boldsymbol\alpha}, \vartheta )$ (\cite
{M1,MM}). This allows one to define the total, elastic and plastic
stress tensors and the thermodynamical forces conjugate to the
parameters $\boldsymbol\alpha$, thus separating different factors in
the dissipation inequality (see 10.21 in \cite{M}, Sec.10.1 or
\cite{M1}).\par

Comparing the expression for the internal energy (\ref{freenergy2})
with these in \cite{M1,MM} one sees that the metric
$\mathbf{g}$ entering the free energy through the
tensor $\mathbf{E}^m$ plays here the role of an internal variable
$\boldsymbol{\alpha }$ and its Ricci tensor $Ric(\mathbf g)$ takes
the place of the space gradient $\nabla \boldsymbol{\alpha}$
(\cite{M1,MM}). The elastic $\mathbf E^{el}$ (respectively inelastic
$\mathbf E^{in}$) strain tensors are direct material analogs of
$\boldsymbol{\epsilon}_{e}$ (respectively of
$\boldsymbol{\epsilon}_{p}$). Thus it is conceivable to adopt the internal variables
approach in searching for the form of the equations
governing the behavior of our system.  \par

We take the Lagrangian in the form
\begin{multline}
L=L(\rho_{ref},\mathbf V,\mathbf E^{el},\mathbf E^{in},\mathbf E^{m},
Ric(\mathbf g),\vartheta, \nabla^{g_{o}}\vartheta )= K-\rho_{ref}u-U(\phi)=\\ =
\frac{\rho_{ref}}{2}\vert \mathbf{V}\vert^{2}_{h}+\rho_{ref}[
\boldsymbol{\gamma}\cdot Ric(\mathbf g)+\mu \Vert
\nabla^{g_{o}}\vartheta \Vert^{2}_{g}+f_{0}(\mathbf E^{in},\mathbf
E^{m},\vartheta )+f(\mathbf E^{in},\mathbf E^{m},\vartheta;\mathbf
E^{el})]-U(\phi)
\end{multline}
 In this expression $\boldsymbol{\gamma}$ is a constitutive tensor, $U$ is the potential of body forces,
 $f_{0}$ is the \emph{"basic inelastic energy"} and $f$ is the strain energy of linear thermoelasticity,
i.e. a quadratic function of the elastic strain tensor with coefficients depending on the
temperature and the remaining inelastic strains:

\beq f(\mathbf E^{in},\mathbf E^{m},\vartheta;\mathbf
E^{el})=[\mathbf{c_{0}}+(\vartheta
-\vartheta_{0})\mathbf{c_{1}}]:\mathbf E^{el}+(\mathbf{e}:\mathbf
E^{el}:\mathbf E^{el}). \eeq \par Here $\mathbf{c_{0}}$ and
$\mathbf{c_{1}}$ are tensors characterizing the interaction of the
elastic processes with the inelastic ones and temperature
respectively (for instance, $\mathbf{c_{1}}$ is the thermal
expansion tensor). The tensor $\mathbf{e}$ is the elasticity
tensor.
\par
Assuming that the decomposition (4.2) is valid, substitution of the
total strain tensor $\mathbf E^{tot}$ instead of the elastic strain
tensor $\mathbf E^{el}$ into the expressions for internal energy and
dissipative potential (below) does not change the form of function (5.3)
\begin{multline}
f(\mathbf E^{in},\mathbf E^{m},\vartheta;\mathbf E^{el})=f(\mathbf E^{in},
\mathbf E^{m},\vartheta;\mathbf E^{tot}-\mathbf E^{in}-\mathbf E^{m})=\\
=[\mathbf{c_{0}}+(\vartheta -\vartheta_{0})\mathbf{c_{1}}]:(\mathbf
E^{tot}-\mathbf E^{in}-\mathbf E^{m}))+
(\mathbf{e}:(\mathbf E^{tot}-\mathbf E^{in}-\mathbf E^{m}):(\mathbf
E^{tot}-\mathbf E^{in}-\mathbf E^{m}))=\\
=+(\mathbf{e}:(\mathbf E^{in}+\mathbf E^{m}):(\mathbf E^{in}+
\mathbf E^{m}))+\\
[(\mathbf{c_{0}}-2\mathbf e:(\mathbf E^{in}+\mathbf
E^{m})+(\vartheta -\vartheta_{0})\mathbf{c_{1}}]:\mathbf
E^{tot}+\mathbf e:\mathbf E^{tot}:\mathbf E^{tot},
\end{multline}
but changes the tensor $\mathbf c_{0}$ and adds a term to the inelastic
energy $f_{0}$.  \par

This allows us to replace $\mathbf E^{el}$ by $\mathbf E^{tot}$ in
the internal energy, so that we can equivalently use Lagrangian in the form
\begin{multline}
L=L[\rho_{ref},\mathbf V,\mathbf E^{tot},\mathbf E^{in},\mathbf E^{m},
Ric(\mathbf g),\vartheta, \nabla^{g_{o}}\vartheta ]= \\ =
\frac{\rho_{ref}}{2}\vert \mathbf{V}\vert^{2}_{h}+ \rho_{ref}[\mathbf{\gamma}\cdot Ric(\mathbf g)+\mu
\Vert \nabla^{g_{o}}\vartheta \Vert^{2}_{g}+f_{0}(\mathbf \mathbf
\mathbf E^{in},\mathbf E^{m},\vartheta )+f(\mathbf E^{in},\mathbf
E^{m},\vartheta;\mathbf E^{tot})]-U(\phi).
\end{multline}
The strain energy here has the form (5.2) where $\mathbf E^{el}$ is replaced by $\mathbf E^{tot}$.\par
 In the dynamic equations it is more convenient to use $\mathbf E^{tot}$ since $\phi$ is a geometrically
 explicit and observable quantity while in the dissipative inequality it is more convenient to
 use $\mathbf E^{el}$ because it allows one to separate inputs of different processes into the entropy production.\par

\subsection{Free energy}
\textbf{Free energy} is defined as usual, by
the equality
\[
\psi=u-s\theta,
\]
where $s$ is the \textbf{specific entropy}.

\subsection{Dissipative potential}
 The \textbf{dissipative (pseudo) potential} id chosen to be a function of the following variables
\beq \mathcal{D}=\mathcal{D}( {\dot {\mathbf E}^{in}},{\dot {\mathbf
E}^{m}};\mathbf E^{m}, \vartheta ). \eeq

We include $\dot {\mathbf E}^{in}$ together with $\dot {\mathbf g}$
to emphasize the difference between the kinetic energy related with
$\phi^{tot}_{t}$ and the inelastic strain rate ${\dot {\mathbf
E}^{in}}$ participating in the process of viscous dissipation, see \cite{M3}.\par

\subsection{Evolution equations}
Introducing the action for the material in a domain $G\subset M$, corresponding to the Lagrangian $L$
\[
A(\phi ,\phi_{1},\mathbf g)=\int_{G}Ldv_{g}
\]
we will write down the equations of evolution for the system
characterized by the dynamical variables $\phi ,\phi_{1},\mathbf{g}$:
\begin{enumerate}
\item Equilibrium equation:
\beq \frac{\delta A}{\delta \phi}=0. \eeq

Since the total deformation $\phi$ enters only through the elastic strain tensor
$\mathbf E^{el}$ this equation is, essentially, the equilibrium
equation of elasticity theory.  If the strain energy is chosen in
the form (5.3), this equation takes the conventional form of the
elasticity equilibrium (Euler) equation with the \emph{elastic moduli depending on the material point and temperature}
 (see (5.3)) \beq   \frac{1}{\g} \frac{\partial (\rho_{ref}\g
h_{ij} V^{j})}{\partial t}-P^{I}_{i\ ;I} = \nu_{i}(\phi(X)). \eeq Here $\boldsymbol\nu=-dU(\phi)$ is the 1-form of the body forces and $P^{I}_{i}$ is the elastic first Piola-Kirchhoff stress tensor (see below Sec.5.5).  Covariant derivative is taken with respect to the material metric $g$.
\item
Equation of plastic deformation:
\beq \frac{\delta A}{\delta \phi^{i}_{1}}=-\frac{\delta \mathbf
D}{\delta {\dot \phi}^{i}_{1}}. \eeq
Notice that in difference to the usual form of this equation (\cite{M1,MM}) we take the variation of dissipative potential by the derivative $\dot{\phi}_{1}$ of \emph{internal variable} $\phi_{1}$ rather then the partial derivative.
 This is necessary due to the fact that $\phi_{1}$ enters Lagrangian through its spacial gradient.\par
To clarify the form of this equation we notice that
\begin{multline}
\frac{\delta A}{\delta \phi^{i}_{1}}=-\frac{1}{\sqrt{\vert g\vert }}\partial_{X^I}\left[ \frac{\partial u}{\partial \phi^{i}_{1,I}}\sqrt{\vert g\vert } \right]=-div_{g}(\mathcal{P}^{I}_{1\ i})=-\left[ T^{in\ N}_{M}\frac{\partial E^{in\ M}_{N}}{\partial \phi^{i}_{1,I}} \right]_{;I}=\\ =-\left[ T^{in\ N}_{M}g^{MK}h_{ij}(\phi^{j}_{,N}\delta^{I}_{K}+\phi^{j}_{K}\delta^{I}_{N}) \right]_{;I}.
\end{multline}
Here $\mathcal{P}^{I}_{1\ i}$ is the first Piola-Kirchhoff stress tensor density of inelastic configuration $\phi_{1}$, see below.  Last equality is due to the fact that
$\frac{\partial E^{in\ M}_{lin\ N}}{\partial \phi^{i}_{1,I}}=g^{MK}h_{ij}(\phi^{j}_{,N}\delta^{I}_{K}+\phi^{j}_{K}\delta^{I}_{N}).$\par

On the other hand,
\[
\frac{\delta \mathbf
D}{\delta {\dot \phi}^{i}_{1}}=-\partial_{K}\left[ \frac{\partial \mathbf{D}}{\partial {\dot E}^{in\ M}_{N}}\frac{\partial \dot{E}^{in\ M}_{N}}{\partial \dot{\phi}^{i}_{1,K}}\right]=-\partial_{K}\left[ \frac{\partial \mathbf{D}}{\partial {\dot E}^{in\ M}_{N}}\cdot \frac{1}{2}h_{ij}(g^{MK}\phi^{i}_{1,N}+g^{MS}\phi^{j}_{1,S}\delta^{K}_{N})\right],
\]
where we have used (4.15).  As a result, equation (5.9) has the form
\beq
div_{g}(\mathcal{P}^{I}_{1\ i})+\partial_{K}\left[ \frac{\partial \mathbf{D}}{\partial {\dot E}^{in\ M}_{N}}\cdot \frac{1}{2}h_{ij}(g^{MK}\phi^{i}_{1,N}+g^{MS}\phi^{j}_{1,S}\delta^{K}_{N})\right]=0.
\eeq
\item Equation of metric evolution: \beq\label{eq3} \frac{\delta A}{\delta
\mathbf{g}}=-\frac{\partial \mathbf D}{\partial  \mathbf{\dot g}}. \eeq
Here we have used the partial derivatives in the right side of equation since $\mathbf{E}^m =\frac{1}{2}ln(\mathbf{g}_{0}^{-1}\mathbf{g})$ depends on $\mathbf{g}$ but not on its derivatives.
\par If the free energy depends on the scalar curvature
$R(\mathbf g)$ instead the full Ricci tensor,
equation (\ref{eq3}) has the form

 \beq{c}\cdot
{\mathcal E}(\mathbf{g})^{IJ}=-\mathcal{T}^{IJ}-\frac{\partial
\mathbf D}{\partial {\dot {\mathbf E}^{m}}}, \eeq
where
\[ \mathcal{T}^{IJ}=\frac{1}{\sqrt{\vert \mathbf{g}\vert}}\frac{\delta}{\delta g_{IJ}}\left( [\frac{\rho_{ref}}{2}\vert
\mathbf{V}\vert^{2}_{h}+\rho_{ref}[\mu \Vert \nabla^{g_{o}}\vartheta
\Vert^{2}_{{g}_{o}}+f_{0}+f]-U(\phi)]\sqrt{\vert\mathbf{
g}\vert}\right)
\]
is the canonical energy-momentum tensor including elastic effects,
effects of inelastic deformation and some thermal effects and
$\mathcal{E}(\mathbf{g})^{IJ}$ is the Einstein tensor of metric $g$
(\cite{MTW}).  If $\mathbf g=\mathbf g_{o}$ is the reference metric
then this equation is absent ($\mathbf{g}_{0}$ is fixed).\par
\end{enumerate}
\begin{remark} In the 2-dim elasticity any metric $g$ in $M$  is Einstein metric, i.e $Ric_{IJ}=\frac{R(g)}{2}g_{IJ}$,\cite{}.
In this case using the scalar curvature $R(g)$ instead of the Ricci tensor in (5.13) does not place any restrictions to the
 material metric $g$.
\end{remark}
\subsection{Stress tensors}

Stress tensors characterizing material response to the
deformations, heating and other physical processes play a crucial
role in the formulation of the evolution equations and dissipative
inequalities. In the material (Lagrangian) formulation there are several
 stress tensors playing different roles in the dynamical picture. They are related to one another
and, through the deformation $\phi$, to the only stress tensor that
is usually present in the Euler picture - the Cauchy tensor
$\boldsymbol\sigma$. Such a plurality of material stress tensors is
related to the presence of two material metrics, i.e. $\mathbf
g_{o}$, $\mathbf g$, used to raise and lower the indices in tensors
and the two different Cauchy metrics - $\mathbf C(\phi),\mathbf C(\phi_{1})$.\par

Here we recall the definitions of the most useful stress tensors through the total internal energy $u$
 or the strain energy $f$ and provide the formulas relating them to one another.

For the total deformation $\phi$ we introduce three stress tensors defined by the differentiation of internal energy by the deformation gradient $F=\phi_{*}$, Cauchy metric $C(\phi)$ and the strain tensor $E^{tot}$.

\begin{table}[h]
\begin{center}
\renewcommand{\arraystretch}{1.25}
\begin{tabular}{| l | c | c | c |}
\hline
Type & I Piola-Kirchhoff & II Piola-Kirchhoff & Strain dual \\ \hline
Tensor & $\mathcal{P}^{I}_{i}=\rho_{ref}\frac{\partial u}{\partial\phi^{i}_{,I}}$ &  $S^{IJ}=2\rho_{ref}\frac{\partial u}{\partial C(\phi)_{IJ}}$ & $T^{el\ I}_{J}=\rho_{ref}\frac{\partial u}{\partial E^{el\ J}_{I}}$\\ \hline
Relations  & $\mathcal{P}^{I}_{i}=J(\phi) \sigma^{j}_{i}\phi^{-1\ I}_{j}$ &  $\mathcal{P}^{Ii}=S^{IK}\phi^{i}_{,K}$ & $T^{el\ I}_{J}=S^{IK}C(\phi_{1})_{KJ}$ \\ \hline
\end{tabular}
\caption{Stress tensors defined by total deformation $\phi$.}
\end{center}
\end{table}
The formula relating $S$ and $T^{el}$ is obtained in the assumption of linear approximation $C(\phi)=C(\phi_{1})+2C(\phi_{1})E^{el}$, see Sec.4.1.
  $J(\phi)$ here is the Jacobian of the total deformation $\phi$ calculated with respect to the metrics $\mathbf h$ and $\mathbf
g_{o}$, see \cite{MH}, Sec.2.2.  Expression for the tensor $S^{tot\ IJ}$  in the Table 1 is the material Doyle-Erickson formula (see \cite{SM}).\par

For the inelastic deformation $\phi_{1}$ we introduce three stress tensors defined by the differentiation of internal energy by the deformation gradient $F=\phi_{1*}$, Cauchy metric $C(\phi_{1})$ and the strain tensor $E^{tin}$ (see Sec.4.1):

\begin{table}[h]
\begin{center}
\renewcommand{\arraystretch}{1.25}
\begin{tabular}{| l | c | c | c |}
\hline
Type & I Piola-Kirchhoff & II Piola-Kirchhoff & Strain dual \\ \hline
Tensor & $\mathcal{P}^{i}_{1\ i}=\rho_{ref}\frac{\partial u}{\partial\phi^{i}_{1,I}}$ &  $S^{IJ}_{1}=2\rho_{ref}\frac{\partial u}{\partial C(\phi)_{1\ IJ}}$ & $T^{in\ I}_{ J}=\rho_{ref}\frac{\partial u}{\partial E^{in\ J}_{I}}$\\ \hline
Relations  & $\mathcal{P}^{I}_{1\ i}=J(\phi_{1}) \sigma^{j}_{i}\phi^{-1\ I}_{j}$ &  $\mathcal{P}^{Ii}_{1}=S^{IK}_{1}\phi^{i}_{1,K}$ & $S^{IK}_{1}g_{KJ}=T^{in\ I}_{ J}$ \\ \hline
\end{tabular}
\caption{Stress tensors defined by inelastic deformation $\phi_{1}$.}
\end{center}
\end{table}
The relation between $S_{1}$ and $T^{in}$ is obtained in the assumption of linear approximation $E^{in}=\frac{1}{2}g^{-1}(C(\phi_{1})-g).$ \par

For the deformation (evolution) of material metric $g_{0}\rightarrow g_{t}$ there are defined the following stress tensors

\begin{table}[h]
\begin{center}
\renewcommand{\arraystretch}{1.25}
\begin{tabular}{| l | c | c | c |}
\hline
Type & Eshelby stress,\cite{EE2,EM,M} & Canonical & Strain dual \\ \hline
Tensor & $b^{i}_{I}=-\rho_{ref}\frac{\partial u}{\partial P^{I}_{i}}$ &  $S^{m\ IJ}=2\rho_{ref}\frac{\partial u}{\partial g_{IJ}}$ & $T^{m\ I}_{J}=\rho_{ref}\frac{\partial u}{\partial E^{m\ J}_{I}}$\\ \hline
Relations  &   &  $b^{i}_{I}=S^{m\ MN}P^{-1\ i}_{M}g_{NI}$ & $S^{m\ IK}g_{0\ KJ}=T^{m\ I}_{ J}$ \\ \hline
\end{tabular}
\caption{Stress tensors defined by deformation of the material metric.}
\end{center}
\end{table}
Here $P^{I}_{X\ i}$ is the uniformity mapping $P:V\rightarrow T_{X}(X)$ and the internal energy is refereed to the reference volume $d_{g_{0}}V$,\cite{EE2}, Ch. 5, Sec.5.5.  Formula relating $S^m$ and $T^{m}$ is obtained in the assumption of linear approximation $E^{m}=\frac{1}{2}g^{-1}_{0}(g-g_0 ).$\par  One can define the variant of the Eshelby stress by  $\tilde{b}^{I}_{J}=-\rho_{0}\frac{\partial u}{\partial D^{J}_{I}}$ using the material (1,1)-tensor $D$.  Its relation to the tensor $b^{i}_{I}$ is given by $\tilde{b}^{I}_{J}=b^{i}_{J}P^{-1\ I}_{0\ i}.$
\par
Notice also that the canonical stress tensor $S^{m}$ is the direct analog of the spacial part of the energy-momentum tensor of the Gravity Theory, \cite{MTW}.\par
\begin{remark}
It is instructive to compare our definition of the \textbf{elastic} first Piola-Kirchhoff tensor with its definition as the difference (comp. \cite{M2}, Ch.10)
\beq T^{el I}_{i}=T^{I}_{i}-\phi_{2,i}^{j}T^{in\ I}_{j}, \eeq
\newline where $\phi_{2}=\phi \circ \phi_{1}^{-1}$ is the elastic part of total deformation.
\end{remark}

\section{Dissipation inequality}
In this section we present the dissipative inequality for the $(\phi, \phi_{1},g)$-model.\par
Below $\nabla$ means $\nabla^{{g}_o}$. We will adopt here
the expression (\ref{freenergy2}) for the free energy but assume,
for simplicity, the quasi-static behavior of the material (i.e. velocity $\mathbf{V}$ is
negligible), potential energy $U$ is absent and internal energy $u$
depends on the scalar curvature $R(\mathbf{g})$ only, instead of on the full Ricci tensor:
\begin{equation}\label{freeen}
\psi=u(\mathbf E^{el},\mathbf E^{in},\mathbf E^m,
R(\mathbf{g}),\vartheta,\nabla^{{g}_o}\vartheta)-s\theta
\end{equation}
\newline
with $\mathbf E^m=\frac 1{2}ln(\mathbf{g}_o^{-1}\mathbf{g})$ and
$\mathbf{g}$ playing the role of an internal parameter
$\boldsymbol\alpha$ (comp. \cite{M}).\par

We will be using the notations: \beq  \tilde s=-\frac{\partial
\psi}{\partial\vartheta},\ \mathbf A=\frac{\partial
\psi}{\partial\nabla\vartheta}, \eeq
\newline
and the formula for the time derivative of the scalar curvature
\begin{equation}
\frac{\partial }{\partial t}R(\mathbf{g})=lim_{\Delta t\rightarrow
0}[R(\mathbf{g})(t+\Delta t)-R(\mathbf{g})(t)]=\boldsymbol{\mathcal
E}(\mathbf{g})\cdot \mathbf{{\dot g}},
\end{equation}
where  $\boldsymbol{\mathcal E}(\mathbf{g})=\frac{\delta
R(\mathbf{g})}{\delta \mathbf{g}}$ is the Einstein tensor of the
material metric $\mathbf{g}$ (see \cite{MTW}).\par
Calculate now the derivative of the free energy:
\begin{multline}\label{derfreeen1}
\dot \psi ={\mathbf T}^{el}\cdot\dot{\mathbf E}^{el}+\mathbf T^{in}\cdot\dot{\mathbf E}^{in}+
\frac{1}{2}\mathbf S^m\cdot \mathbf{\dot g}+\boldsymbol{\mathcal E}(\mathbf{g})\cdot \mathbf{{\dot g}}-\\
-\tilde s\cdot\dot\theta+\nabla\cdot(\mathbf A\dot\vartheta)-(\nabla\cdot\mathbf A)\dot\vartheta,
\end{multline}
\newline
where  the vectorial identity $\nabla\cdot(\mathbf A\dot\vartheta)=(\nabla\cdot\mathbf A)\dot\vartheta+
\mathbf A\cdot\dot{\nabla\vartheta }$ was used and where tensors $T^{in},T^{el},S^m$ are as in the Sec.5.\par

Recalling the formula for the variation $\frac{\delta \psi}{\delta \mathbf{g}
}=\frac{\partial \psi }{\partial \mathbf{g}}-\nabla\cdot\frac{\partial \psi
}{\partial\nabla^{\mathbf{g}_o} \mathbf{g}}$ and using the notation $\mathcal
A=\frac{\delta \psi}{\delta \mathbf{g} }$ we find
\begin{equation}
[\frac{1}{2}\mathbf S^m +\boldsymbol{\mathcal E}(\mathbf{g})]\cdot\mathbf{\dot
g} =\frac{\delta \psi}{\delta \mathbf{g}}\cdot\mathbf{\dot
g}=\mathcal A(\mathbf g)\cdot\mathbf{\dot g}.
\end{equation}
 Then the time derivative of the free energy takes the form
\begin{equation}\label{derfreeen2}
\dot \psi ={\mathbf T}^{el}\cdot\dot{\mathbf E}^{el}+\mathbf
T^{in}\cdot\dot{\mathbf E}^{in}+\mathcal A(\mathbf{g})\cdot \mathbf{{\dot g}}
-\tilde s\cdot\dot\vartheta-\nabla\cdot(\mathbf
A)\dot\vartheta+\nabla\cdot(\mathbf A\dot\vartheta ).
\end{equation}
\par Recall now the Gibbs inequality for a thermodynamical system with
internal parameter $\boldsymbol\alpha$ (here $\boldsymbol\alpha
=\mathbf{g}$), see \cite{M1},
\begin{equation}\label{Gibbs}
-(\dot \psi +s\dot\theta)+p_{i}+\nabla\cdot(\theta\mathbf
k)-(s\cdot\nabla)\theta\geq 0.
\end{equation}
Here \[p_{i}={\mathbf T}\cdot\dot{\mathbf E}^{tot}\]
 is the power of the internal work, stress tensor $\mathbf T$ will be specified below and $\mathbf k$ is the extra entropy flux density assumed
 to include contributions from the flux of the internal variables. \par
Substituting the expression (\ref{derfreeen2})  for $\dot \psi $ into the
Gibbs inequality (\ref{Gibbs}) we get
\begin{eqnarray}
&&-{\mathbf T}^{el}\cdot\dot{\mathbf E}^{el}-\mathbf T^{in}\cdot\dot{\mathbf E}^{in}-\mathcal
A(g)\cdot\mathbf{{\dot g}}
+\tilde s\cdot\dot\vartheta+\nabla\cdot(\mathbf A\dot\vartheta)-\\
&&\nabla\cdot(\mathbf A\dot\vartheta-s\dot\theta) +p_{i} +\nabla\cdot(\vartheta\mathbf
k)-(s\cdot\nabla)\vartheta\geq 0
\end{eqnarray}\newline
 In the special case of when one uses the linearized definitions of strain tensors (see Sec. 4.1) and the commutativity condition that allows to write the total strain rate $\dot{E}^{tot}$ in the form (4.2) is fulfilled, the previous inequality takes the form
\begin{multline}
({\mathbf T}(1+2E^{m})(1+2E^{in})-{\mathbf T}^{el})\cdot\dot{\mathbf E}^{el}+({\mathbf T}(1+2E^{m})(1+2E^{el})-\mathbf T^{in})\cdot\dot{\mathbf
E}^{in}+\\+\mathbf T (1+2E^{el})(1+2E^{in})\cdot\dot{\mathbf E}^{m}
-\mathcal A(\mathbf g)\cdot\mathbf{\dot g}\\
 +(s -\tilde s+\nabla\cdot\mathbf A)\dot\vartheta+\nabla\cdot(\theta\mathbf k-\mathbf A\dot\theta)
-(s\cdot\nabla)\theta\geq 0.
\end{multline} \par
Now we use the fact that the derivatives ${\dot E}^{el},\dot \vartheta$ are controllable variables and can take arbitrary positive and negative values and therefore, their coefficients should be equal zero, (\cite{M1}).  Thus we obtain the relations
\begin{equation}
\mathbf T={\mathbf T}^{el}(1+2E^{in})^{-1}(1+2E^{m})^{-1},
\end{equation}
and
\begin{equation}
s=-(\tilde s-\nabla\cdot\mathbf A)=-\Big(\frac{\partial
\psi}{\partial\vartheta}-\nabla\cdot\frac{\partial \psi
}{\partial\nabla\vartheta}\Big) =-\frac{\delta
\psi}{\delta\vartheta}
\end{equation} \par
Assuming for $\mathbf k$ the prescription
\begin{equation}
\mathbf k=\vartheta^{-1}\mathbf A\dot\vartheta=
\vartheta^{-1}\frac{\partial
\psi}{\partial\nabla\vartheta}\dot\vartheta ,
\end{equation}
\newline
the reduced dissipation inequality is obtained in the form:
\begin{multline}\label{dissineq3}
[{\mathbf T}^{el}(1+2E^{in})^{-1}(1+2E^{el})-\mathbf T^{in}]\cdot\dot{\mathbf
E}^{in}+\\ +
[{\mathbf T}^{el}(1+2E^{in})^{-1}(1+2E^{m})^{-1}(1+2E^{el})(1+2E^{in})-2\mathcal A(\mathbf
g)g_{0}]\cdot\dot{\mathbf E}^{m}
-(s\cdot\nabla)\vartheta   \geq 0
\end{multline}
\newline
where we have used the expression $\dot{\mathbf E}^{m}= \frac{1}{2}\dot{(g^{-1}_{0}g-I)}=\frac{1}{2}g^{-1}_{0}\dot{g}$ for the linearized metric strain tensor $E^{m}= \frac{1}{2}(g^{-1}_{0}g-I)$.\par

Dissipation inequality  (6.14) is satisfied if one request the independent fulfillment of the stronger conditions - two intrinsic dissipation inequalities
\begin{equation}\label{dissineq4}
\begin{cases}
[{\mathbf T}^{el}(1+2E^{in})^{-1}(1+2E^{el})-\mathbf T^{in}]\cdot\dot{\mathbf
E}^{in}  \geq 0 ,\\
[T^{el}(1+2E^{in})^{-1}(1+2E^{m})^{-1}(1+2E^{el})(1+2E^{in})-2\mathcal A(\mathbf
g)g_{0}]\cdot\dot{\mathbf E}^{m} \geq 0.
\end{cases}
\end{equation}
\newline
and the thermal dissipation inequality
\begin{equation}\label{dissineq5}
-(s\cdot\nabla)\theta \geq 0.
\end{equation}
Using the relation between the tensor $S^m$ and $T^m$ presented in the Table 3 we can rewrite second inequality in the form
\beq
[T^{el}(1+2E^{in})^{-1}(1+2E^{m})^{-1}(1+2E^{el})(1+2E^{in})-T^{m}-2\mathcal{E}(g)g_{0}]\cdot\dot{\mathbf E}^{m} \geq 0
\eeq
Comparing inequalities (6.15-6.17) with similar dissipative inequalities in \cite{M1,} we see that the coefficient of $\dot{E}^{in}$ (respectively $\dot{E}^{m}$) can be interpreted as the \emph{effective stress tensor} for integrable inelastic deformation (respectively, for evolution of the uniform structure). Such modifications of the stress tensors are customary in studying of the entropy production by a combination of interrelated elastic and inelastic processes, comp. \cite{M}, Ch.10.\par
\subsection{Yield condition from dissipative inequality}
If all three strain tensors in (6.15) are small (in comparison with the unit tensor), the inequalities (6.15) take the form
\beq
\begin{cases}
[{\mathbf T}^{el}-\mathbf T^{in}]\cdot\dot{\mathbf
E}^{in}  \geqq 0 ,\\
[T^{el}-T^{m}-2\mathcal{E}(g)g_{0}]\cdot \dot{\mathbf E}^{m} \geqq 0
\end{cases}
\eeq
These inequalities can be interpreted as the yield conditions determining when the corresponding type of inelastic evolution (plastic integrable: $\dot{\phi}_{1}\ne 0$ and/or material metric $\dot{g}_{t}\ne 0$  respectively) may proceed.  In each case the elastic stress $\mathbf{T}^{el}$ should be large enough to overcome the barrier necessary for initiation of the corresponding process. \par
This form of yield condition is similar to the condition for the plastic deformation to proceed obtained from the Drucker postulate, see \cite{SC}, Sec.8.11, inequality (8.85).\par  Solutions of evolution equations (5.9) and (5.11) describes also the evolution of stress tensors $T^{in},T^{m}$.  Therefore the conditions (6.18) for elastic stress $T^{el}$ evolves in time.  This evolution can be related to the \emph{hardening processes} during an elasto-plastic deformation of materials.  \par
Consider, for instance a homogeneous isotropic case.  Let $Q_{IJ}$ be a symmetric (0,2)-tensor. The evolution in the direction of this tensor, i.e. the evolution for which $\dot{E}^{in}_{IJ}=\lambda(t)Q_{IJ},\ \lambda(t)>0$ may proceed only if the difference $(T^{el}_{IJ}-T^{in}_{IJ})$ is such that $Tr_{g}((T^{el}_{IJ}-T^{in}_{IJ})Q^{IJ})\geqq 0$, i.e. if this difference is positive in the direction of tensor $Q_{IJ}$.\par
Leaving further study and comparison of these conditions with the usual yield criteria, \cite{SC,FG} for future work, we notice only that the conditions (6.18) are anisotropic by its nature and might possibly provide useful supplement to the usual criteria in essentially anisotropic situations.

\section{Conclusions}

In this work we analyzed the relation between the Bilby-Kroner -Lee
multiplicative decomposition $\mathbf F=\mathbf F^{e}\mathbf F^{p}$
of the total deformation gradient into elastic and plastic factors
(\cite{M2,EM,Kr,L}) and the theory of uniform materials
(\cite{TW,N,W}).  We prove that the Bilby-Kroner-Lee multiplicative decomposition
is equivalent to the uniform material model with two deformation
mappings, i.e. the total $\phi$  and the inelastic $\phi_{1}$
deformations together with the uniformity structure. Uniformity enters
through the (1,1)-tensor field $\mathbf D$ in the material manifold $M$
or through the material metric $\mathbf g$. We introduced
 the total, the elastic and the inelastic strain tensors characterizing
different types of the geometrical evolution of the material. After
discussing the relations between these strain tensors and the
deformation gradients $\mathbf F^{e}$ and $\mathbf F^{p}$, we chose
the form of the internal energy (5.2) and of the
dissipative potential (5.6) for the materials modeled by the triple $(\phi
,\phi_{1},\mathbf g)$.  The evolution equations were written down
for all dynamical variables $(\phi ,\phi_{1},\mathbf g)$.  We
discussed different types of stress tensors that naturally enter the
scheme of our work. Finally we wrote down the dissipative
inequalities for the materials of $(\phi ,\phi_{1},\mathbf g)$-type
where the terms corresponding to the different types of dissipative
processes are separated.\par

Further research along the lines indicated in this paper seems to be in order. First, one should compare
our results with those obtained by G.Maugin in a different framework \cite{M,M1}. Second, in the continuation of
this work we will study the evolution equations (5.7-5.12), obtain the energy balance law and the heat equation that
follows from it along the lines of \cite{M}. Third, some special cases and examples will be considered.
 \vskip0.4cm \par
\textbf{Acknowledgements}. The authors would like to thank professor M.Elzanowski who made the manuscript of the book \cite{EE2}
available to them before the publication and for valuable advices during the preparation of the manuscript.  We also
\section{Appendix} In this Appendix we present calculation of the total strain rate $\dot E^{tot}$ that was used in Sec.6.\par
From the formula (4.2) for the linearized definition of strain tensors we get
\[
E^{tot}=g_{0}^{-1}C(\phi_{1})E^{el}+g_{0}^{-1}gE^{in}+E^{m}.
\]
Taking derivative we get
\beq
\dot{E}^{tot}=g_{0}^{-1}\dot{C(\phi_{1})}E^{el}+g_{0}^{-1}C(\phi_{1})\dot{E}^{el}+ g_{0}^{-1}\dot{g}E^{in}+g_{0}^{-1}g\dot{E}^{in}+\dot{E}^{m}.
\eeq
From the definition of linearized  $E^{in}=\frac{1}{2}g^{-1}(C(\phi_{1})-g)$ we get $C(\phi_{1})=g+2gE^{in}$ and, therefore, $\dot{C}(\phi_{1})=\dot{g}+2\dot{g}E^{in}+2g\dot{E}^{in}.$ As a result, \[g_{0}^{-1}\dot{C(\phi_{1})}=g_{0}^{-1}(\dot{g}+2\dot{g}E^{in}+2g\dot{E}^{in})=2\dot{E}^{m}+4\dot{E}^{m}E^{in}+2(1+2E^{m})\dot{E}^{in},\]
where we have used $g_{0}^{-1}g=1+2E^{m}$.  \par
In the second term in (8.1) $g_{0}^{-1}C(\phi_{1})=g_{0}^{-1}gg^{-1}C(\phi_{1})=(1+2E^{m})(1+2E^{in})$, in the third one $g_{0}^{-1}\dot{g}=2\dot{E}^{m}$.  Substituting these expressions into (8.1) and collecting coefficients of strain rate tensors we get
\beq
\dot{E}^{tot}=(1+2E^{m})(1+2E^{in})\dot{E}^{el}+(1+2E^{m})\dot{E}^{in}(1+2E^{el})+\dot{E}^{m}(1+2E^{el})(1+2E^{in}).
\eeq
In a case where strain tensors participating in the second and third terms of the last formula commute with the corresponding strain rate tensor, we get
\beq
\dot{E}^{tot}=(1+2E^{m})(1+2E^{in})\dot{E}^{el}+(1+2E^{m})(1+2E^{el})\dot{E}^{in}+(1+2E^{el})(1+2E^{in})\dot{E}^{m}
\eeq

\end{document}